\documentclass[preprint,letter,preprintnumbers,nofootinbib,amsmath,amssymb,floatfix,superscriptaddress]{revtex4}

\usepackage{graphicx}
\usepackage{dcolumn}
\usepackage{bm}
\usepackage{slashed}
\usepackage{kevin}
\usepackage{tabularx}

\begin{document}
\title{Radiative energy loss and $v_2$ spectra for viscous hydrodynamics}
\author{Kevin Dusling}
\affiliation{Physics Department, Building 510A\\Brookhaven National Laboratory\\Upton, NY-11973, United States}
\author{Guy D.~Moore}
\affiliation
    {%
    Department of Physics,
    McGill University,
    3600 rue University,
    Montr\'{e}al, QC H3A 2T8, Canada
    }%
\author{Derek Teaney}
\affiliation{Department of Physics and Astronomy, Stony Brook University,
Stony Brook, New York 11794-3800, United States}
\date{\today}

\begin{abstract}
This work investigates 
the first correction to the equilibrium phase
space distribution and its effects on spectra and elliptic flow in heavy ion collisions.
We show that the
departure from equilibrium on the freezeout surface is the largest part of the viscous
corrections to $v_2(p_T)$.  However, the momentum dependence of the
departure from equilibrium is not known {\it a priori}, and it is probably not
proportional to $p_T^2$ as has been assumed in hydrodynamic simulations. At high
momentum in 
weakly coupled plasmas it is determined by the rate of radiative energy loss
and is proportional to $p_T^{3/2}$.  The weaker $p_T$ dependence
leads to straighter $v_2(p_T)$ curves at the same value of
viscosity.  Further, the departure from equilibrium is generally species
dependent. A species dependent equilibration rate, with baryons equilibrating
faster than mesons, can explain ``constituent quark scaling'' without invoking
coalescence models.
\end{abstract}

\maketitle

\section{Introduction}

When two ultra-relativistic nuclei collide, they leave behind a region
of high energy-density QCD matter, whose properties we would like to
understand better.  The initial geometry of the QCD matter is set by the
overlap region of the two colliding nuclei. 
Generally, the nuclei 
collide at finite impact parameter rather than head-on.
In this 
case the initial geometry is not a disk, but is an
``almond shaped'' ellipse.  (The short and long axis
of the initial almond are taken as the $x$ and $y$ axes respectively.)
The production mechanism of
the QCD matter is local and knows nothing of this global geometry.
Therefore, to a first approximation the initial stress tensor will be locally 
azimuthally symmetric. 
Subsequently, if there are no reinteractions the produced matter
will free stream to the detector; 
the initial
geometry will have no influence on the evolution,
and the angular distribution of the final observed hadrons 
will also be azimuthally symmetric.
On the other hand,  if there are strong interactions which
maintain local thermal equilibrium,
the pressure gradients in the $x$ direction
will be  larger than in the $y$ direction, an  
 anisotropy in the collective flow will develop,  and  ultimately
an anisotropy in the momentum spectrum  of the final hadrons
will be observed.

The final momentum anisotropy is characterized experimentally by $v_2$,
the second harmonic of the azimuthal distribution of the produced particles
with respect to the reaction plane. 
Experimentalists have measured $v_2$
as a function of transverse momentum $p_T$,
particle type, and impact parameter
\cite{Adams:2005dq,Adcox:2004mh,Back:2004je,Arsene:2004fa}.  
These results are surprisingly
well described by ideal hydrodynamics \cite{idealhydro}, which amounts
to the approximation that the interactions are fast enough to maintain the
matter in equilibrium from an early time until hadronic freeze-out.
There are some limits to this success.  First, the measured
$v_2$ falls below the ideal hydrodynamic prediction for momenta larger than  $p_T \gsim 2.0\,{\rm GeV}$.
Second, the hydro fit fails to reproduce certain relative trends observed in 
the baryon and meson elliptic flows. These trends are compactly summarized
by ``constituent quark scaling" \cite{Abelev:2008ed,Adare:2006ti,coalesence} which generally has been attributed
to  a kind of coalescence of constituent quarks \cite{Lin:2001zk,Molnar:2003ff,Greco:2003xt,Fries:2003vb}. Here we
will argue that the first corrections to equilibrium can clarify
both of these shortcomings without the need for a coalescence model.

To quantify the corrections to ideal hydrodynamics it is important to
study nonideal (viscous) hydrodynamics.  In the last two years there has
been a major push in this direction
\cite{Baier:2006gy,Romatschke:2007jx,Romatschke:2007mq,Song:2007fn,Dusling:2007gi,Huovinen:2008te,Song:2007ux,Bozek:2007qt}.  These studies have used various formalisms and
have studied variations of $v_2(p_T)$ with respect to the input shear
viscosity, the model for the initial geometry, and various other nuisance
parameters.  However, we want to point out here that these studies have all made a
common assumption about the way that the asymmetry in the stress tensor
is manifested in the particle distribution after freezeout.  In
particular, the particle distribution after freezeout is locally of the form
$f = f_0 + \delta f$ where $f_0$ is the 
equilibrium distribution and $\delta f$ is the 
first correction. All groups have assumed that
$\delta f(p) \propto p^2 f_0$ and that the coefficient
of proportionality is independent of particle type.

In this paper we will argue that this
assumption matters, and that it is far from secure.  After an overview
of the issue in the next section, in \Sect{sec:examples} we will discuss the physics which
establishes the momentum dependence of $\delta f$ and its behavior in
several theories.  We will see that
while the most studied theories give $\delta f \propto p^2 f_0$, the
most QCD-like theories do not.  Then we explore the behavior of
multi-component plasmas in \Sect{sec:multicomponent}.  We see
there that the viscous corrections $\delta f(p)$ for different species
are generically different.  This fact can account for the ``constituent quark
scaling'' observed in the baryon
and meson elliptic flows without any reference to the hadronization 
process.
We then make our concluding remarks. Some technical material is postponed to
appendices.

Throughout, we will denote 4-vectors with capital letters $P,Q$ and use $\p,\q$
for their 3-vector components, $E_p,E_q$ for their energy components, and $p,q$
for $|\p|,|\q|$.  Our metric convention is [--,+,+,+],  so that $u_{\mu}
u^{\mu} = -1$.  We use tilde to indicate momenta scaled by temperature,
$\tilde{p} \equiv p/T$.  We will mostly  write $n_p$  for the equilibrium
distribution function, $n_p \equiv 1/(\exp(p/T) \mp 1)$ but will occasionally use
$f_0(p)$  when common convention dictates its use.  The appropriate statistics will be clear
from context.

\section{Overview}

The energy momentum tensor is given by the sum of its ideal and
dissipative parts%
\footnote{We use Landau-Lifshitz conventions to fix $\epsilon,u^\mu$ in
  terms of four components of $T^{\mu\nu}$.  The other six independent
  components of $T^{\mu\nu}$ can always be accommodated by a
  $\pi^{\mu\nu}$ satisfying $u_\mu \pi^{\mu\nu}=0$.}
\beqa
T^{\mu\nu}=(\epsilon+\mathcal{P})u^\mu u^\nu + \mathcal{P}g^{\mu\nu}+\pi^{\mu\nu},
\eeqa
and obeys the equation of motion,
\beqa
\partial_\mu T^{\mu\nu}=0 \, .
\label{eq:EOM}
\eeqa
In the first-order (or Navier-Stokes) approximation the dissipative part
of the stress energy tensor in the local rest frame is
\beqa
\pi^{ij}=-\eta\left(\partial^i u^j+\partial^j u^i -
\frac{2}{3}\delta^{ij}\partial_k u^k\right) \equiv -\eta\sigma^{ij}
 \equiv-2\eta \langle \partial^i u^{j}\rangle,
\label{eq:evol}
\eeqa
where $\eta$ is the shear viscosity, and we use
$\langle \ldots \rangle$ to indicate that the bracketed tensor
should be symmetrized and made traceless.  It is well known
that the first order theory is
plagued with difficulties such as causality violations and
instabilities \cite{Hiscock:1983zz,Hiscock:1985zz}.  In order to circumvent these issues a second order
theory is required.  The most commonly used second order relativistic viscous
hydrodynamics is due to Israel and Stewart
\cite{IS}.  For technical reasons we use a theory developed by
{\"O}ttinger and Grmela \cite{OG,Ottinger}.  The two theories are
qualitatively the same ({\em i.e.} for sufficiently small
relaxation times they both approach the first order theory).
To streamline the presentation we postpone the details of our
hydrodynamic model to \app{App:details} and refer to previous 
work \cite{Dusling:2007gi}.

The solutions to the hydrodynamic equations yield the underlying temperature and flow profiles in the presence of viscosity.  Particle spectra are then computed using the Cooper-Frye \cite{CF} formula
\beqa
E\frac{d^3N}{d^3p}=\frac{\nu}{(2\pi)^3}\int_\sigma f(\tilde{p}) p^\mu d\sigma_\mu,
\eeqa
where $\tilde{p}\equiv p/T$ and $\sigma_\mu$ is the freeze-out hypersurface taken as a surface of constant energy density in this work.  For a system out of equilibrium $f(\tilde{p})$ is not the equilibrium distribution function but also contains viscous corrections,
\beqa
f(\tilde{p})=f_0(\tilde{p})+\delta f(\tilde{p}),
\eeqa
where $f_0$ is the ideal Bose/Fermi distribution function.  The form of
$\delta f$ is constrained by the requirement that $T^{ij}$ be continuous
across the freeze-out hypersurface:
\beqa
T^{\mu\nu} = \nu \int \frac{d^3 p}{(2\pi)^3 p^0} p^\mu p^\nu f(\tilde{p})
\qquad \rightarrow \qquad
\pi^{ij} = \nu \int \frac{d^3 p}{(2\pi)^3 p^0} p^i p^j \delta
f(\tilde{p}) \,.
\label{deltaf_and_pi}
\eeqa
Dropping $\delta f$ from the
final particle spectra is inconsistent as it leads to a discontinuity in
$T^{\mu\nu}$.  The  form for
$\delta f$ which satisfies continuity  in the
local rest frame is proportional to $\hat{p}^i\hat{p}^j \pi_{ij}$  and is traditionally parametrized by $\chi(p)$\;\footnote{In actual simulations $\pi^{ij}$ is
treated as a dynamical variable in a second order fluid formalism. Then
to first order one can make the replacement,
$\llangle \partial_i u_j \rrangle  \rightarrow -\pi_{ij}/2\eta$. There has been
no attempt to systematically include $\delta f$ through second order in hydrodynamic simulations.}
\beqa
\label{chip}
\delta f(\p)&=&-n_p(1\pm n_p) \chi(\p) \, , \\
            &=&-\chi(p) n_p (1 \pm n_p)
\hat{p}^i \hat{p}^j \llangle \partial_i u_j \rrangle \, ,
\eeqa
where we have distinguished $\chi(p)$ and  $\chi(\p) \equiv \chi(p)
\hat{p}^i \hat{p}^j \llangle \partial_i u_j \rrangle$ by the
argument of the function.
One moment of $\chi(p)$ is fixed by the shear viscosity (see below)
but otherwise $\chi(p)$ is an arbitrary function of $p$.  To date all works on
viscous hydrodynamics have taken the quadratic \Ansatz
~and have usually worked in a Boltzmann approximation
\beqa
\chi(p)\propto p^2 \, .
\eeqa

\begin{figure}
\includegraphics[scale=1]{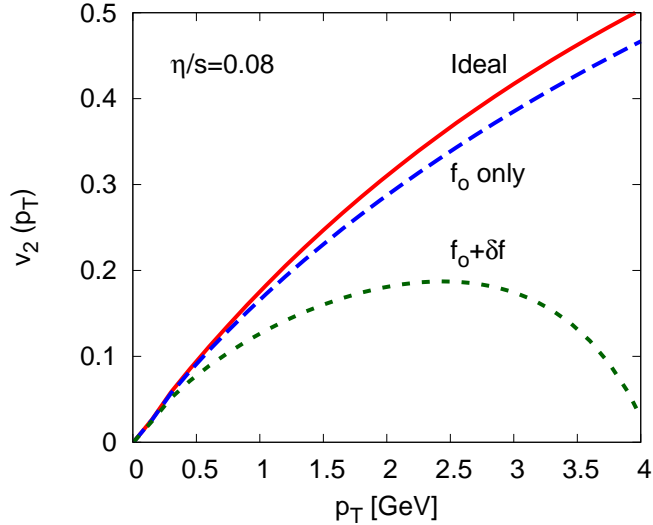}
\caption{Typical results for $v_2(p_T)$ from a viscous hydrodynamic model employing the quadratic \Ansatz.  The run parameters are $\eta/s=0.08$, $T_{\scriptscriptstyle \rm frzout}=140$ MeV and $p=1/3\epsilon$. Further details are in \app{App:details}.
\label{fig:quad}}
\end{figure}

Let us look at typical results for $v_2(p_T)$ as shown in
\Fig{fig:quad}.  The curve labeled `Ideal' shows the result using
ideal hydrodynamics ({\em i.e.~}$\eta/s=10^{-6})$.  The curve labeled
`$f_0$' shows the resulting elliptic flow from the viscous evolution
(the solution of Eqs.~(\ref{eq:EOM}),(\ref{eq:evol}) and (\ref{eq:evol2}))
but
without including the viscous correction to the distribution function.
In other words, this shows how the viscous correction to the temperature
and flow profiles manifests itself in the particle spectra.  Only modest
corrections to the spectra are found.  As already emphasized, this
result is unphysical since dropping $\delta f$ violates continuity of
the stress tensor.  Last, the curve labeled `$f_0+\delta f$' also takes
$\delta f$ into account, using the quadratic \Ansatz.  The viscous
correction to the distribution function dominates the reduction in $v_2$
at large $p_T$.  That means that the $\delta f$ term is responsible for
a significant part  of the effects of viscosity in the particle spectra.

 This being the case, it is imperative to perform a systematic study on the
form of the viscous correction as well as its effect on elliptic flow.  Most of
this paper will discuss the form of the viscous correction appearing in
weakly-coupled QCD.  Although this is not a theory of hadronizing QCD, it is
one theory where quantitative first-principle calculations can be performed.
One of our major findings is that not all models of energy loss give the same
predictions for the off-equilibrium distribution function.

\section{Form of $\delta f$ in several theories}
\label{sec:examples}

In this section we consider a number of theories, to show that while
the dependence $\chi(p) \propto p^2$ is expected in some cases,
other functional dependence is
expected in others,  including weakly coupled QCD and a hadron
(resonance) gas.
The theories where we can make a definite statement about the functional
form of $\delta f$ are all described by kinetic theory.  Since
freeze-out is defined as the point where scatterings go from being
common to being rare on the time scale of the evolution of the system,
we generally expect that, {\em just} before freezeout, kinetic theory
should be a reasonable description.

Within kinetic theory, the
distribution function $f(\p,\x)$ is determined by a Boltzmann equation,
\be
\label{Boltzmann}
(\partial_t + v_{\bf p}\cdot \partial_{\bf x} ) f(\p,\x) = -{\rm C}[f,\p]\, ,
\ee
where ${\rm C}[f,\p]$ is the collision operator.
In equilibrium the distribution function obeys
\be
   n(\p, \x)  =  \frac{1}{e^{-P_{\mu} u^{\mu}(t,\x)/T(t,x)}  \mp 1} \, ,
\qquad
\mbox{with} \qquad   {\rm C} [n,\p] = 0 \, .
\ee
To determine the first viscous correction $\delta f$, we work in a vicinity of
the local rest frame $u^{\mu} = (1, u^{i}(\x,t))$, and substitute  $f = n(\p,\x)
+  \delta f$ into \Eq{Boltzmann} keeping terms first order in the spatial
derivatives
\be
\label{linboltz}
   \frac{p^{i}{p^j}}{E_p T} n_p(1 \pm n_p) \llangle \partial_i u_j \rrangle
 = -{\cal C}[\delta f,\p] \, .
\ee
Here ${\cal C}[\delta f, \p]$ denotes the linearized collision operator, $i.e.$ the collision operator expanded to first order in $\delta f$.
In writing \Eq{linboltz} we have
used  ideal hydrodynamics and thermodynamic relations
to rewrite time derivatives as spatial derivatives,  and we have
neglected  gradients proportional  $\partial_i u^{i} $
which are responsible for the bulk viscosity \cite{Teaney:2009qa}.  \Eq{linboltz} is an integral equation
for $\delta f$ which can be solved by various methods.

Since the first viscous correction is a scalar and
must be proportional to the the strains, the most general form for the viscous
correction in the local rest frame  can be parametrized  by the
function $\chi(p)$ as in \Eq{chip}.
Close to equilibrium  the first viscous correction
$\delta f$ determines the strains
\be
 \pi^{ij} = - 2 \eta \llangle  \partial^i u^{j} \rrangle = \int \frac{d^3\p}{(2\pi)^3} \frac{p^{i} p^{j}}{E_\p}\, \delta f \, ,
\ee
which ultimately yields a relation between between the
shear viscosity and the viscous correction $\chi(p)$
\be
\label{constraint}
   \eta = \frac{1}{15} \int \frac{d^3\p}{(2\pi)^3} \frac{p^2}{E_\p}  n_p(1 \pm
n_p) \chi(p) \, .
\ee
This is the only general constraint on the functional form of the viscous
distribution function.  To proceed further we must specify completely the
form of the linearized collision operator, which we will do in the
context of various model theories.

\subsection{Simplest model:  relaxation time approximation}

The simplest model (really a cartoon) for the collision operator is the
relaxation time approximation,
\beqa
{\rm C}[\delta f,\p] = \frac{f(\p) - f_0(\p)}{\tau_R(E_p)} \, ,
\eeqa
where $\tau_R$ is the momentum dependent relaxation time to be specified.
Substituting this form for the collision operator into \Eq{linboltz},
and working in a Boltzmann approximation $n_p(1\pm n_p) \rightarrow n_p$
yields the following form for $\delta f$:
\beqa
\label{deltaf}
\delta f = -\frac{\tau_R(E_\p)}{T E_\p}n_p p^ip^j\langle \partial_i u_j\rangle \, .
\eeqa

Note however that the relaxation time is in general energy dependent.
In different theories, $\tau_R(E_p)$ might show different functional
dependence on $E_p$.  Without details about the dynamics of the theory
in question, we can only
parametrize the viscous correction.
Here we will discuss a massless classical gas where $n_p = e^{-p/T}$
and parameterize the relaxation time (or the distribution function) with
a simple power law
\beqa
\delta f(p)&=&-n_p\chi(\tilde{p})\hat{p}^i \hat{p}^j
  \left<\partial_i u_j\right>,\nonumber\\
\chi(\tilde{p})&=&C(\alpha)\tilde{p}^{2-\alpha}\,.
\label{deltaf2}
\eeqa
The constant, $C(\alpha)$, is determined through \Eq{constraint}:
\beqa
C(\alpha) = \frac{120\eta}{(\epsilon+\mathcal{P})\Gamma(6-\alpha)} \,.
\eeqa

There are two limiting cases for the functional form of the
the relaxation time approximation, $\alpha=0$ and $\alpha=1$.
The momentum dependence of the relaxation time in these extreme cases
is
\beqa
\tau_R(p)  \propto \left\{ \begin{array}{rl}
 p &  \qquad \alpha=0 \mbox{~} \mbox{ (quadratic ansatz),} \\
\mbox{const} & \qquad  \alpha=1 \mbox{~} \mbox{ (linear ansatz).} \end{array}\right.
\eeqa
Most theories will lie between these  two extreme
limits.\footnote{There are exceptions to this rule. For instance, in a gas of Goldstone bosons far below
  the symmetry breaking scale one expects $\alpha=2$, since the
cross section grows rapidly with energy,  $\sigma \sim  E^2/\Lambda^4$. }  Loosely speaking,
if the energy loss of
high momentum particles grows  linearly with momentum, $\frac{dp}{dt} \propto p$ one
expects a relaxation time independent of momentum, $\tau_R \propto p^0$.
On the other hand if the energy loss approaches a constant $\frac{dp}{dt} \propto \mbox{const}$,  the relaxation time will grow with the particle momentum $\tau_R \propto p$.

\begin{figure}
\begin{center}
\includegraphics[width=0.49\textwidth]{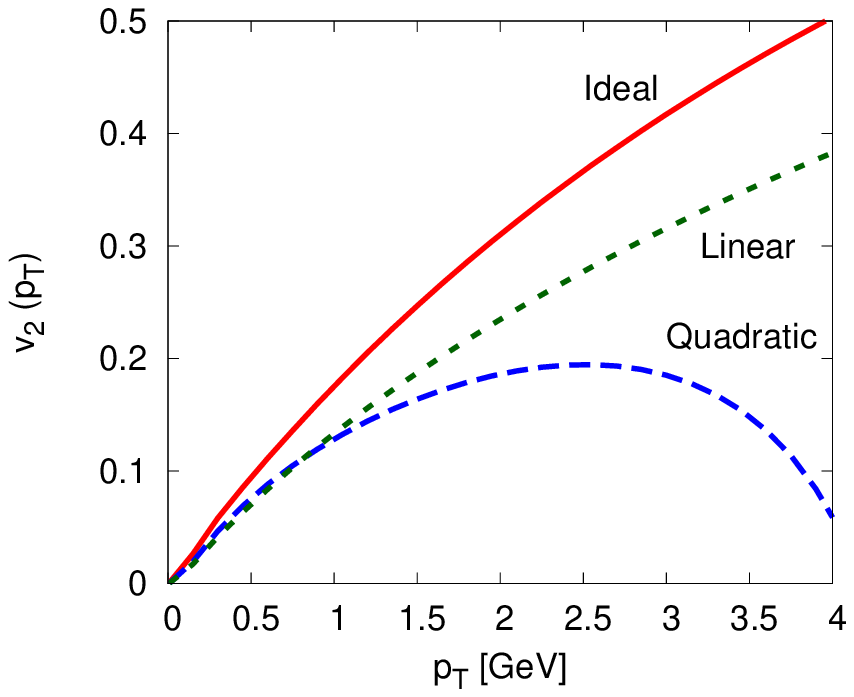}
\includegraphics[width=0.49\textwidth]{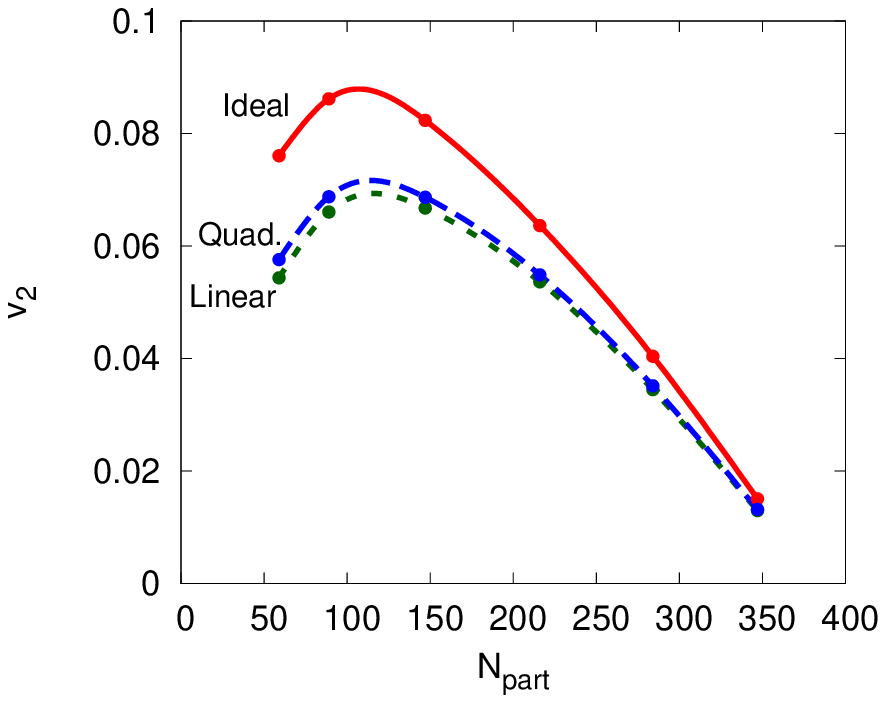}
\end{center}
\caption{Left: $v_2(p_T)$ using the Linear or
  Quadratic \Ansatze\ for the distribution function.  Right: Integrated
  $v_2$ versus centrality showing independence from the precise form of
  the viscous correction.  Run parameters can be found in \Fig{fig:quad}.}
\label{fig:v2quadlinear}
\end{figure}

\Fig{fig:v2quadlinear} shows the elliptic flow computed using these two
functional forms for the first viscous correction. It is
important to emphasize that  shear
viscosity is the same in both cases. Examining these
figures, we see that the integrated elliptic flow is largely
insensitive to the functional form of the first viscous correction.
This is because the integrated $v_2$ is primarily
determined by the hydrodynamic variables $e,u^{\mu}$,$\pi^{\mu\nu}$
which are independent of the functional dependence of the relaxation
time \cite{Teaney:2009qa}.
The  differential elliptic flow $v_2(p_T)$ is sensitive to the rate
of equilibration especially above $p_T \simeq 1.2 \, {\rm GeV}$.

\subsection{Scalar $\lambda\phi^4$ theory}

Scalar field theory has been described at length by Jeon
\cite{Jeon:1994if}, who rigorously derived the Boltzmann equation and
its collision kernel and then solved for $\chi(\tilde{p})$ numerically.
But if we make the approximation of Boltzmann statistics, we can
actually solve for $\delta f$ in closed form.

First the non-linear Boltzmann equation with Bose-Einstein statistics
is
\be
C[f,\p]= \int_{\kk,\p',\kk'} \Gamma_{\p \kk\rightarrow \p'\kk'}
\left[ f_\p f_\kk (1 + f_{\p'})(1 + f_{\kk'}) - f_{\p'} f_{\kk'}(1 + f_\p)(1 + f_\kk) \right] \,,
\ee
where the transition rate (including a final state symmetry factor) is
\be
\label{eq:Gamma}
\Gamma_{\p\kk\rightarrow \p'\kk'}
= \frac{1}{2}\, \frac{\left| \mathcal M \right|^2 }{(2 E_\p) (2 E_\kk) (2 E_{\p'}) (2E_{\kk'}) }  (2\pi)^4
\delta^4(P + K - P' - K') \, ,
\ee
and we have used the traditional short hand, $\int_\p = \int \frac{d^3\p}{(2\pi)^3}$.
After linearizing,
$f(\p) \rightarrow n_p + n_p(1 \pm n_p) \chi(\p)$ the
linearized collision integral is
\be
\label{22cofdf}
\frac{p^i p^j}{T E_p } \llangle \partial_i u_j \rrangle =
\int_{\kk,\p',\kk'} \Gamma_{\p \kk\rightarrow \p'\kk'}\;  n_p n_k (1 + n_{p'})(1+n_{k'})\,  \left[ \chi(\p) + \chi(\kk) - \chi(\p') - \chi(\kk')\right] \, .
\ee
At this point we will make the Boltzmann assumption by
neglecting the stimulation factors, $(1 + n_p) \rightarrow 1$, and
using,  $n_p = e^{-p/T}$.
Then we will try a solution of the form
\be
\label{eq:phiquad}
\chi(\p) = C p^{i} p^j \llangle \partial_i u_j \rrangle/T^3\, ,
\ee
$i.e.$  assuming  that $\chi(\tilde p) \propto \tilde p^2$ or $\alpha=0$.
Substituting
this form into the integral equation (\Eq{linboltz} and \Eq{22cofdf}), and performing
the integrals yields (see \app{collision})
\beqa
\label{scalar_res}
\llangle \partial_i u_j \rrangle \frac{p_i p_j}{T}
 & = &
C \llangle \partial_i u_j  \rrangle p_i p_j  \frac{\lambda^2}{384\pi^3 T}\,.
\eeqa
Thus,  taking
$C=384 \pi^3/\lambda^2$, the quadratic form in
\Eq{eq:phiquad}
has provided an exact solution solution to
the linearized integral equation.  The viscosity is
$\eta = 1536 \pi T^3/\lambda^2$ in a Boltzmann approximation.

Physically, this happens because of the form of the scattering
cross-section.  Since $\sigma \propto \lambda^2/s$ and $s \propto p$,
the cross-section scales as the inverse of the particle's energy.  The
typical scattering is nearly randomizing, but high energy particles
undergo fewer scatterings than low-energy ones.  Therefore we find the
same functional form as for momentum diffusion but for very different
reasons.

This example from scalar field theory, together with the example of
momentum diffusion (see \Sect{momentum_diffuse} below), are the reason
that most people assume the quadratic ansatz, $\chi(p) \propto p^2$
should hold.

\subsection{Weakly coupled pure-glue QCD}

In this section we will use the Boltzmann equation for pure-glue
QCD in three approximation
schemes to calculate the first viscous correction.
First we will consider a leading $\log(T/m_D)$ approximation
where the dynamics can be summarized by a Fokker-Planck equation
which describes the momentum diffusion of quasi-particles.
In this  limit we will find that the viscous correction is
quadratic at large momentum, $\chi(p) \propto p^2$.
Next we will consider the QCD Boltzmann equation  but
consider only $2 \rightarrow 2$ collisions  and neglect collinear
radiation. In this limit,  we will find that
the viscous correction at large momentum behaves as
$\chi(p) \propto p^2/\log(p)$.
Finally, we will also include collinear radiation in the
Boltzmann  equation as is necessary in a
complete leading order treatment \cite{Arnold:2001ba,Arnold:2002zm}.
We will find that collinear radiation controls the relaxation of the high momentum
modes and asymptotically we have $\chi(p) \propto p^{3/2}$, where the
coefficient of proportionality is set by the rate of
transverse momentum broadening, $\hat{q}$.
The impatient reader may skip to \Sect{qcdsum} which summarizes
the results of these three approximation schemes.

\subsubsection{Momentum diffusion in a leading log treatment}
\label{momentum_diffuse}
In a leading  log  approximation,
$\log(T/m_{D})$ is  considered a  large number
and
the dynamics describes soft Coulomb scattering. Each soft collision
involves a small momentum transfer of order $q\sim g T$, but these
collisions happen relatively frequently at a rate of $\sim g^2 T$ (neglecting logarithms).
Thus a typical particle with momentum $T$ will diffuse in momentum
space and equilibrate on a time scale of $\sim g^4 T$. The resulting
Boltzmann equation linearized around equilibrium can be written as a Fokker-Planck equation \cite{Arnold:2000dr,Juhee}
\be
\label{leading_log}
\partial_t \delta f + v_\p \cdot \partial_x \delta f =
T\mu \frac{\partial}{\partial p^i}\left(  n_p(1 + n_p) \frac{\partial} {\partial p^i} \left[ \frac{\delta f(\p) }{n_p (1 + n_p)} \right] \right) + \mbox{gain terms}  \, ,
\ee
where  $\mu$ is the drag coefficient of a high momentum  gluon  in 
this approximation scheme\,\cite{Thoma:1992kq,Braaten:1991jj}
\be
\label{bthoma_soft}
\frac{d\p}{dt} = -\mu \hat{\p}\,,  \qquad \mbox{with} \qquad \mu =  \frac{g^4C_A^2}{24 \pi} T^2 \log\left(\frac{T}{m_D}\right)   \, .
\ee
The precise form of the gain terms has been given in \cite{Arnold:2000dr,Juhee},
but only involves the $\ell=0,1$  spherical harmonic components of $\delta f(\p)$, $i.e.$ \, $\int d\Omega_\p \, \delta f(\p)$ and $\int d\Omega_\p \, \hat{\p} \, \delta f(\p)$ .
In the hydrodynamic limit considered here
$\delta f(\p)$  is proportional to a traceless rank 2 tensor  ($\hat p^i \hat p^j - \delta^{ij}/3$) and  these gain terms  vanish. Substituting
the form of \Eq{chip} into \Eq{leading_log} leads to the following equation
for $\chi(p)$:
\be
\label{ll_fokker}
 n_p (1 + n_p) \frac{p }{T}  =  T\mu n_p (1 + n_p)
 \left(-\frac{d^2}{dp^2} + \left(\frac{1 + 2n_p}{T} - \frac{2}{p}
 \right) \frac{d}{dp} +  \frac{6}{p^2} \right) \chi(p) \, .
\ee

We are not aware of a closed form solution to this equation, but
we can find a solution for $\chi(p)$ at large momentum.
Making the approximation $1+2n_p \approx 1$, we find that
\be
\chi(p)=\frac{p^2}{2 T \mu }
\ee
solves this equation.  This is the well known {\em quadratic} \Ansatz.

\subsubsection{Boltzmann equation with $2\rightarrow 2$ collisions }

We  next will consider the QCD Boltzmann equation but we will neglect
collinear radiation. We emphasize that this is not a consistent approximation
scheme. Nevertheless, it illustrates clearly the relative roles
of hard collisions and inelastic processes in determining the
functional form of $\chi(p)$ in the relevant sub-asymptotic regime.

The linearized Boltzmann equation is the same as
\Eq{22cofdf},
but the squared matrix element is
\be
\label{megg}
\left| \mathcal M \right|^2=
8g^4 C_A^2 \left(3-\frac{ut}{s^2}-\frac{us}{t^2}-\frac{ts}{u^2}\right) \, ,
\ee
which describes $2\rightarrow 2$ gluon scattering  after
summing over all spins and colors and dividing by the gluon degeneracy
factor $2d_A$.
These matrix elements must be dynamically screened
using Hard Thermal Loops. A procedure which is consistent at leading
order (where the Debye mass is small) but which makes a reasonable
estimate when the Debye mass is not small has also been described in
\cite{Arnold:2003zc}, and we can follow exactly the numerical procedure
of that reference to find $\chi(\tilde p)$.%
\footnote{Some minor technical difficulties are discussed in the next section. }
We can also study the asymptotic behavior more directly.
At asymptotically large momentum where $\log(\tilde p)$ may
be considered large,  \app{collision} shows that
\be
 \chi(p) \propto  \frac{p^2}{\log(p/T) }\,.
\ee
The constant in front of the log is related to
$\llangle dE/dt\rrangle_p$,
the rate of collisional energy loss of
a gluon with momentum $p$,
\be
\label{chi_coll}
\chi(p) = \frac{p^2}{2T\llangle dE/dt \rrangle_p}\,.
\ee
In
a leading $\ln(p/T)$ approximation the loss rate is \cite{Bjorken:1982tu,Braaten:1991jj}
\be
\label{bthoma}
\left< \frac{dE}{dt}\right>_p = \frac{g^4 C_A^2}{48\pi} T^2 \log\left(\frac{p}{T}\right)\,,
\ee
as is rederived in \app{collision}.
The above asymptotic form agrees well with the numerical solution of the
Boltzmann equation.

\subsubsection{A leading order treatment at asymptotically large
  momenta}
\label{LO}

Early calculations of the shear viscosity in pure-glue QCD found
$\chi(\tilde p) \propto \tilde p^2$, that is,
$\alpha=0$ \cite{Baym:1990uj,Heiselberg:1994vy}.
However this is because they were
leading-log treatments, which reduced to momentum diffusion
discussed above.  It was
realized in \cite{Arnold:2001ba,Arnold:2002zm} that inelastic number changing processes
are only suppressed by a log, but are enhanced at large energy $E$ by a
factor of $(E/T)^{1/2}$ and dominate equilibration for
$E/T > \log(1/g)$.%

This should not be a surprise.  After all, if we think about
``equilibration'' (energy loss) in  QED,  
we find that although the leading order 
mechanism for the energy loss of a high
energy electron is ionization (elastic scattering),
 bremsstrahlung actually dominates the loss rate. This is 
the case because in bremsstrahlung the
energy lost per scattering can scale with the incident energy, rather than 
being incident energy independent as is the case with ionization.  As a result, the penetration depth of an
electromagnetic shower scales only logarithmically with the incident
energy, {\it i.e.} the relaxation time is constant up to logs,
 $\tau_R \propto E^0$.
If the same behavior occurred in QCD we would expect the linear \Ansatz\
to hold, $\alpha=1$.

The current understanding of energy loss in perturbative QCD is that 
the high energy behavior lies
between these extremes.  High-energy  particles in a QCD
plasma lose energy predominantly by inelastic gluon radiation and the
time scale for energy loss is short compared to the time scale for
momentum diffusion (``jet broadening'').  In particular it was shown by
Baier {\it et al} that for $E\gg T$ the rate of (inelastic)
energy loss scales with the incident energy as
$dE/dt \propto E^{1/2}$, with the half-integer power arising from the LPM
suppression \cite{Baier:1996kr,Baier:1996sk}.  This implies a ``relaxation
time'' which scales as $\tau_R \sim E / (dE/dt) \propto E^{1/2}$, and therefore
$\alpha =1/2$ \cite{Arnold:2003zc}.  Let us see how this
emerges in the behavior of pure-glue QCD.

The point is that the Boltzmann equation for a gluon plasma possesses
both an elastic scattering term and an inelastic effective $1\rightarrow
2$ scattering term,
\beqa
\partial_t f+v_{\bf p}\cdot \partial_{\bf x}
f=-\mathcal{C}^{2\leftrightarrow 2}[f]
-\mathcal{C}^{1\leftrightarrow 2}[f] \, .
\label{eq:bltz1}
\eeqa
This equation was first solved
at leading order in $\alpha_s$ by Arnold, Moore and
Yaffe to determine the shear viscosity \cite{Arnold:2003zc}.
Their approach involved writing a multi-parameter \Ansatz\ for
$\chi(\tilde p)$ in terms of a basis of test functions.  While the
determination of $\eta$ improves quadratically with the test function
basis, the determination of $\chi(\tilde p)$ improves only linearly.
Therefore to get good accuracy out to $p=15T$ requires the use of a 
large basis of functions. We find a basis of eight functions is sufficient and
the $\tilde p^{3/2}$ behavior is already clear with such a basis.%
\footnote{In fact we find greatly improved convergence of the
  large-momentum behavior, both in terms of basis set size and numerical
  integration precision, by changing the test functions of
  \cite{Arnold:2003zc} to a set which show the correct large momentum
  asymptotic behavior by multiplying $\phi_{2\ldots N}$ defined in
  Eq.(2.32) of the reference by $\tilde{p}^{-1/2}$.}

We can also directly establish the asymptotic form of the solution.  At
asymptotically high momentum near collinear \brem dominates the
equilibration of gluons.  We therefore look at the Boltzmann equation
including only $1\rightarrow 2$ splittings,
\be
\partial_t f+v_{\bf p}\cdot \partial_{\bf x}
f=-\mathcal{C}^{1\rightarrow 2}[f] \, .
\ee
The relevant collision integral for near collinear joining and splitting
of gluons at leading order in $\alpha_s$ was worked out in
\cite{Arnold:2002ja}:
\be
\mathcal{C}^{1\to 2}=\frac{(2\pi)^3}{2\vert{\bf p}\vert^2
  \nu_g}\int_0^\infty dp\pr dk\pr \delta(\vert{\bf
  p}\vert-p\pr-k\pr)\gamma({\bf p};p\pr,k\pr)\left[f_\p(1+f_{\p\pr})(1+f_{\kk})
            -f_{\p\pr}f_{\kk\pr}(1+f_{\p})\right] \, ,
\ee
and is given in terms of the splitting function for $g\to gg$.  In
general this splitting function involves the solution of
an integral equation which includes the LPM effect.  However, in the
{\em deep} LPM regime \cite{Arnold:2008zu} where $\ln^{-1}(\tilde{p})$
can be treated as small, the following leading log result for the
splitting function can be obtained,
\be
\gamma^g_{gg}(p;xp,(1-x)p)=\frac{4\alpha_s C_A d_A}{(2\pi)^4}
   \sqrt{3p\hat{q}}\frac{\left[1-x(1-x)\right]^{5/2}}
   {\left[x(1-x)\right]^{3/2}} \, .
\ee
The above splitting function contains the transport parameter $\hat{q}$, which
characterizes the typical transverse momentum squared transferred to the
particle per unit length.  With the above splitting function we show in
\app{app:qhat} that the solution of the off-equilibrium distribution function
at asymptotically large momentum is
\beqa
\chi_g(p)\approx\frac{0.7}{\alpha_s T \sqrt{\hat{q}}}p^{3/2} \, .
\eeqa

\subsubsection{Summary of weakly coupled pure glue QCD}
\label{qcdsum}
\begin{figure}
\includegraphics[width=0.55\textwidth]{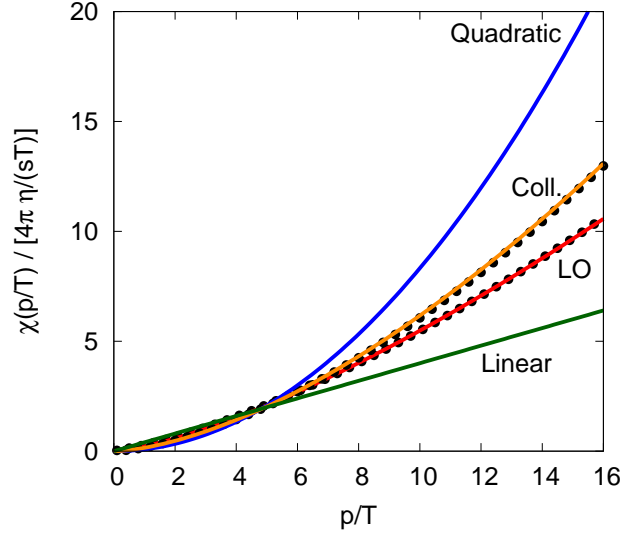}
\caption{The points are from the numerical solution of the Boltzmann equation
for pure glue at leading order (LO) and without $1\leftrightarrow 2$ processes
({\em i.e.} Collisional energy loss only).  The lines are $\chi\propto
p^{2-\alpha}$ for $\alpha=2,1.6,1.38,1$ going from top to bottom.}
\label{fig:summary}
\end{figure}

Let us now summarize some of the main features of the off-equilibrium dynamics
of pure glue QCD at weak coupling.  In the previous three sections we looked at
the behavior of the off-equilibrium correction for pure glue QCD in various
approximation schemes, deriving asymptotic behavior in each case.  These
asymptotics are listed in \Tab{eloss}.  In this section we wish to focus on the
phenomenologically more interesting region where the equilibrating parton has
intermediate energies ($p\sim 10T$).  In this case one must resort to
numerical solutions of the Boltzmann equation which we present in
\Fig{fig:summary}.

To summarize \Fig{fig:summary}, we will discuss the curves
from top to bottom starting with  the ``Quadratic" curve.
In the leading $\log(T/m_D)$ approximation the linearized Boltzmann equation simplifies to
a differential equation, \Eq{ll_fokker}.  The numerical solution to this has been worked
out in \cite{Heiselberg:1994vy,Arnold:2000dr,Juhee} and is well described for all momenta by the asymptotic
quadratic form, $\chi=p^2/2T\mu$.  The numerical result
will be presented in a forthcoming work \cite{Juhee}, and for now we show the
quadratic result as the solid blue line (color online).
Next we considered QCD with the $2\to 2$ gluon scattering matrix element at
leading order.  The agreement between the asymptotics derived in the previous
section and the numerical solution can be found in \app{collision}.  At
intermediate momentum we show the numerical solution of
the Boltzmann equation without inelastic processes as the data points under the
curve labeled ``Coll.''.  The solid curve is the result of a power law fit at
intermediate momentum, $\chi\propto p^{1.6}$.  In the leading order (LO) treatment
when \brem is included, we find further equilibration of the gluons and our
numerical results are reasonably described by the fit,
$\chi\propto p^{1.38}$.
Finally, the linear
ansatz is also shown in \Fig{fig:summary} for comparison.

\begin{figure}
\begin{center}
\includegraphics[width=0.55\textwidth]{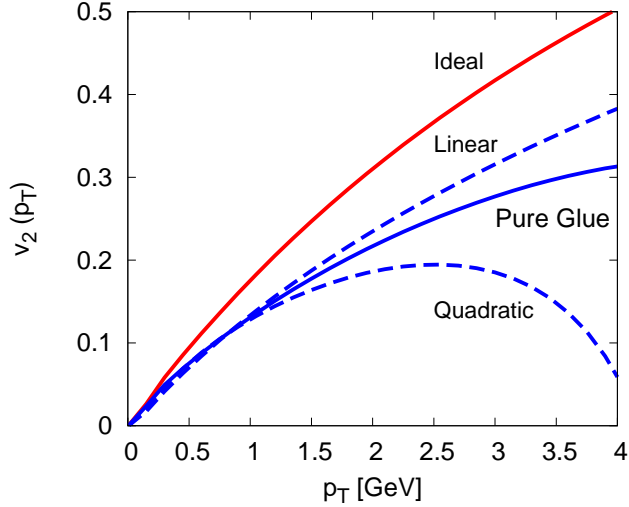}
\end{center}
\caption{$v_2(p_T)$ for a perturbative gluon gas at leading order.  The linear
and quadratic \Ansatze\ are shown for comparison.  Run parameters can be found in \Fig{fig:quad}.}
\label{fig:v2pureglue}
\end{figure}

One can now ask how the observed $\chi\propto p^{1.38}$ of pure glue at leading
order will affect the viscous corrections to elliptic flow.  First of all, as
we have already shown, the integrated $v_2$ will change marginally.  The
differential $v_2$, on the other hand, will be largely affected at higher
$p_T$.  This result is shown in \Fig{fig:v2pureglue} along with the quadratic
and linear \Ansatze~for comparison.

\begin{figure}
\begin{center}
\includegraphics[width=0.55\textwidth]{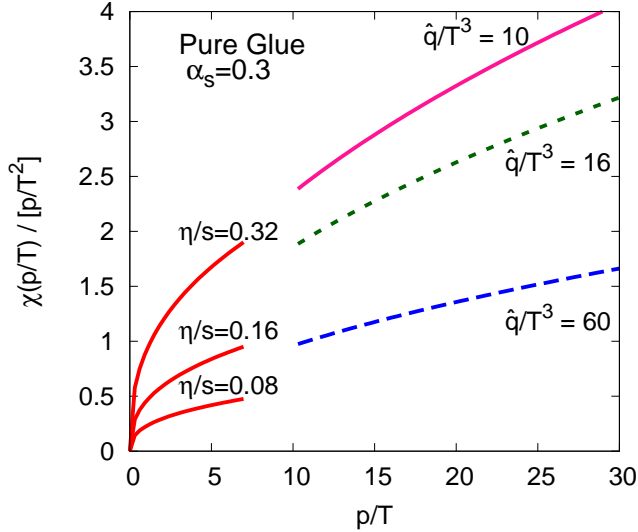}
\end{center}
\caption{The curves at {\em lower} momentum are $\chi(p/T)$ for a perturbative
gluon gas at leading order for three values of the non-perturbative parameter
$\eta/s$.  The curves at {\em higher} momentum show the asymptotic forms of
$\chi$ for three values of the non-perturbative parameter $\hat{q}$. There must
be a consistency between $\eta/s$ and $\hat{q}$ in order that the curves merge
at intermediate momentum.}
\label{fig:eloss}
\end{figure}

The above considerations have shown that the relaxation of the high energy tail
of the distribution is largely controlled by energy loss.  The low /
intermediate momentum region is constrained by the shear viscosity via
\Eq{constraint}.  The {\em strength} of the off equilibrium correction is
controlled by two non-perturbative parameters: $\eta$ at low momentum and
$\hat{q}$ at high momentum.  This is clearly seen by looking at the forms of
$\chi$ we have found for pure glue QCD at leading order,
\beqa
\chi(p)=  \left\{ \begin{array}{rl}
 \frac{2.84 \eta}{sT}\tilde{p}^{1.38} &  \qquad 5\lesssim \tilde p \lesssim 10 \\
\frac{0.7}{\alpha_s T\sqrt{\hat{q}}} p^{1.5} & \qquad   \ln^{-1}(\tilde{p})\ll 1 \end{array}\right. \, .
\eeqa
In \Fig{fig:eloss} we show plots of $\chi$ for various choices of the
non-perturbative parameters $\eta/s$ and $\hat{q}/T^3$.  The main point to take
away is the need for a consistency between $\eta$ and $\hat{q}$ such that the
low and high momentum regions of $\chi$ can merge smoothly into one another.
The three values of $\hat{q}/T^3 = 10,16,60$ we have chosen reproduce the
experimentally observed $R_{AA}$ \cite{Bass:2008rv} when convoluted with the
Higher Twist \cite{HT}, AMY \cite{AMYeloss,Arnold:2001ba,Arnold:2000dr} and ASW
\cite{Baier:1996sk,Baier:1996kr,ASW} energy loss models respectively.  It
appears to be difficult to reconcile the discontinuity of $\chi$ between the
lowest shear viscosity $\eta/s=0.08$ and smallest value of $\hat{q}$ used in
modeling heavy ion collisions.

\subsection{Hadron gas}

One might also ask what scattering behavior is expected at lower
temperatures, in a hadron gas.  How do the highest energy hadrons
equilibrate, as a function of hadron energy?
A complete study requires understanding the energy-dependent
hadron-hadron cross section, which has nontrivial energy dependence and
must be determined from experiment.  However we should be able to say
something about the high momentum behavior.

In hadron-hadron scattering, the inelastic branching fraction rises with
increasing $s$, dominating the cross-section for kinetic energies well
above $\Lambda_{_{\rm QCD}}$.  Since generically no daughter in an
inelastic collision carries more than half the energy of the initial
high $p$ particle, we can take scatterings to be
momentum randomizing (the relaxation time approximation is sensible),
especially for the highest energy hadrons.  The relaxation time is then
controlled by the scattering rate, $\tau_R \sim n \sigma$, with $n$ the
hadron number density and $\sigma$ an averaged total hadronic
cross-section.  So what is the behavior of
the total hadronic cross-section?  At low momenta it is complicated by
resonances but at large momenta there is universally a rising total
cross-section.  Therefore the relaxation time $\tau_R(E)$ should naively
involve a small or zero power of $E$, that is,
$\alpha \sim 1$ is expected, at least for the very high energy tail.%
\footnote{Froissart behavior $\sigma \propto \ln^2(s)$ suggests
$\tau_R \propto \ln^{-2}(p)$.}
Certainly we do not expect $\alpha=0$.  However any more detailed
discussion must be either model or data driven and lies outside the
scope of this paper.

\section{Multi-component plasmas}
\label{sec:multicomponent}

The plasmas just considered are treated as single-component, in the sense
that all degrees of freedom are related to each other by symmetries
(spins by parity, colors by gauge invariance).
The quark-gluon plasma is a multi-component plasma.  Treating $m_s$ as
small and $m_c$ as large, the three light quark types behave the same,
but the gluons behave differently from the quarks.
Similarly, the hadronic plasma present at lower temperatures contains
both baryons and mesons, each of several types.  The different
components generically have different departures from equilibrium, that
is, $\chi_{quark} \neq \chi_{gluon}$, which would manifest as different viscous
corrections to their $p_T$ spectra.  In particular, we will argue that
faster equilibration for baryons than for mesons can give a simple
explanation for the ``constituent quark scaling'' \cite{Abelev:2008ed,Adare:2006ti,coalesence} observed in
$v_2(p_T)$ for mesons and baryons, without invoking any model of
coalescence.

\subsection{Quark-Gluon plasma}

We now consider a two component gas of quarks and gluons and label the
distribution functions with subscripts $q$ and $g$ respectively:
\beqa
\delta f_g(p)&=&-n_p(1+n_p)\chi_g(\tilde{p})\hat{p}^i \hat{p}^j\left< \partial_i u_j\right>\,,\nonumber\\
\delta f_q(p)&=&-n_p(1-n_p)\chi_q(\tilde{p})\hat{p}^i \hat{p}^j\left<
\partial_i u_j\right> \,.
\eeqa
For use in hydrodynamic simulations we will again fit the
off-equilibrium component of the quarks' and gluons' distribution
function to the following power law,
\beqa
\chi_g(\tilde{p})&=&C_g(\alpha_g)\tilde{p}^{2-\alpha_g}\,,\nonumber\\
\chi_q(\tilde{p})&=&C_q(\alpha_q)\tilde{p}^{2-\alpha_q}\,.
\eeqa

\begin{figure}
\begin{center}
\includegraphics[scale=1]{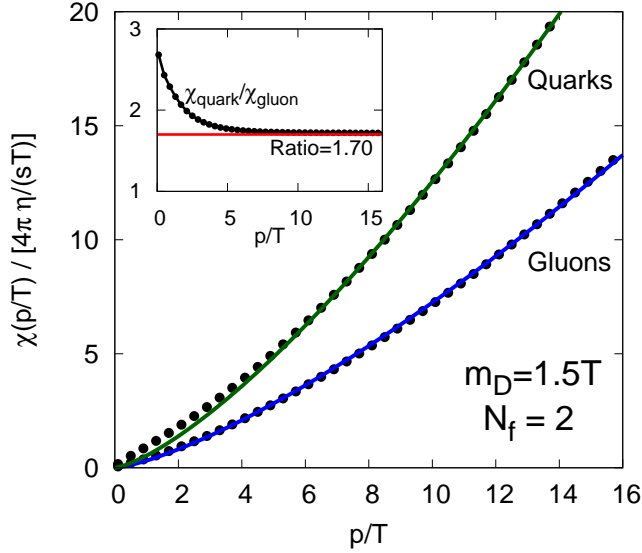}
\end{center}
\caption{Off-equilibrium correction for the case of a perturbative two-flavor
QGP evaluated at leading order.  The sub-figure shows the ratio of the quark to
gluon correction which asymptotically approaches
$\chi_{quark}/\chi_{gluon}\approx 1.7$.
\label{fig:chi2comp}
}
\end{figure}

The results of the numerical solution of the Boltzmann equation for the
two component case are shown as points in \Fig{fig:chi2comp}.  The
solid curves are the results of the fit done at intermediate momentum
($5\leq \tilde{p} \leq 15$) with the result $\alpha_q\approx \alpha_g\approx 0.62$.

In order to solve for the two constants ($C_q$ and $C_g$) we need two constrains.
The first constraint relates the coefficients $C_{q,g}$ to the shear viscosity,
\beqa
\eta = \frac{1}{15}\sum_{a=q,g}\nu_a C_a \int\frac{d^3p}{(2\pi)^3} p^{3-\alpha_a} n_p\left(1\pm n_p\right).
\eeqa
The sum is over quarks and gluons with degeneracies $\nu_g=2d_A =16$ and $\nu_q=4d_f N_f=24$.
The second constraint comes from fixing the ratio of $\chi_q/\chi_g$ to
the numerical solution of the Boltzmann equation.  This ratio is shown
 in \Fig{fig:chi2comp} and at large enough momentum
($\tilde{p}\gtrsim 5$) we find
\beqa
\frac{\chi_q}{\chi_g}\approx 1.70 \,.
\eeqa
The explicit computation of the two coefficients ($C_{q,g}$) in terms of the above ratio and $\eta/s$ is worked out in \app{app:twocomp}.

\begin{figure}
\includegraphics[width=0.49\textwidth]{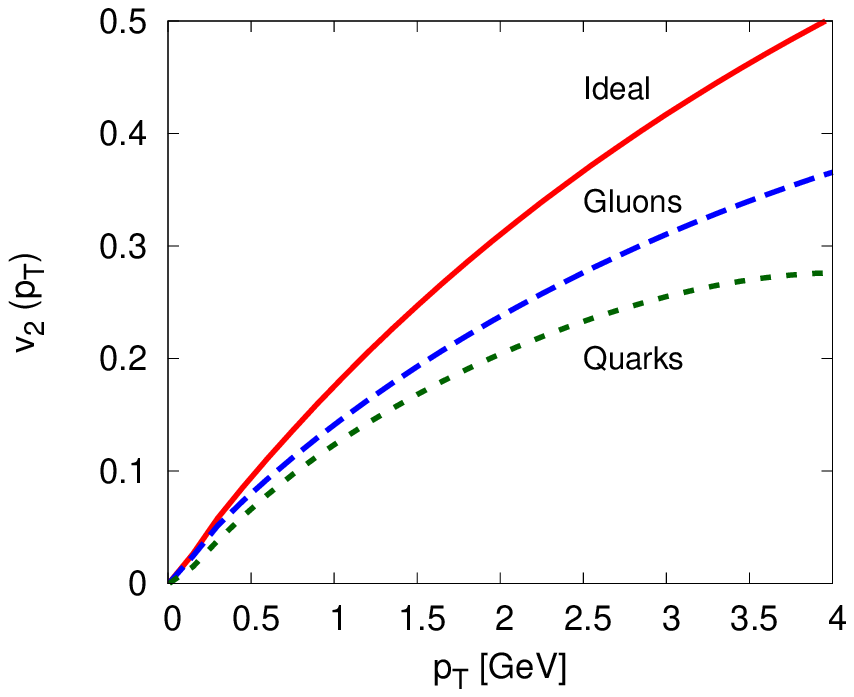}
\includegraphics[width=0.49\textwidth]{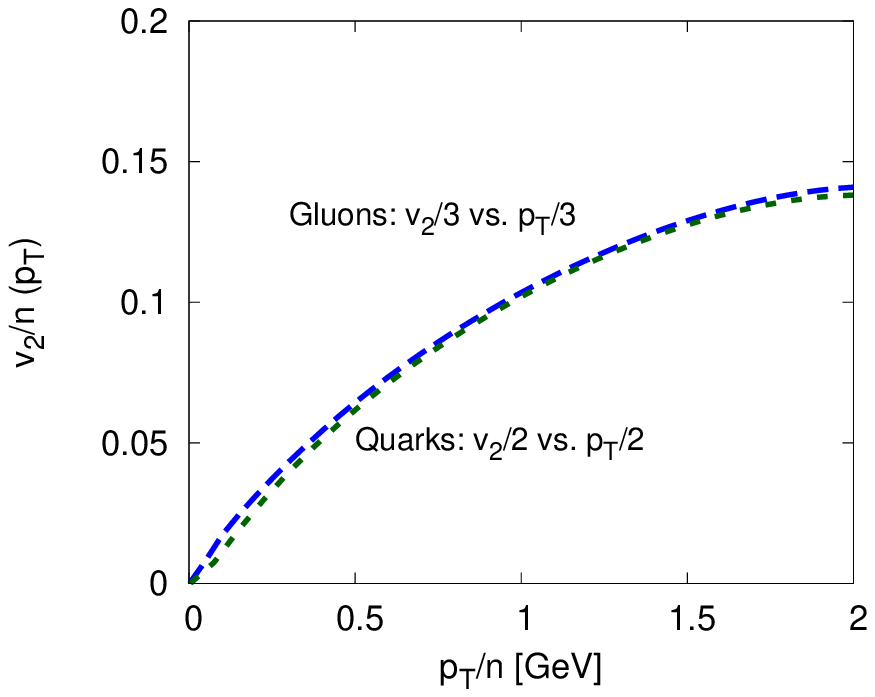}
\caption{Left: Elliptic flow of quarks and gluons.  Right: Both $v_2$ and $p_T$ scaled by n=3,2 for gluons and quarks respectively. Run parameters can be found in \Fig{fig:quad}.}
\label{fig:v2qgp}
\end{figure}

In \Fig{fig:v2qgp} we show the elliptic flow of quarks and
gluons.  Note the larger suppression for quarks as the gluons are forced
into equilibrium much quicker.  This quicker relaxation can not simply
be explained by naively assuming Casimir scaling, $\chi_q/\chi_g\approx
C_A/C_F=2.25$.  Instead this ratio involves a playoff between the faster
equilibration rate of gluons and the tendency of identity changing
processes $q\bar{q}\leftrightarrow gg$, $\QQG$, $\GQQ$ to equilibrate
disequilibrium between the quarks and gluons.
This ratio is evaluated analytically at asymptotically large momentum in
\app{app:qhat}.

The distinct quark and gluon elliptic flow is completely due to the different viscous corrections, which in turn is related to the different relaxation rates of quarks and gluons.  Let us note that if we scale both the $v_2$ and $p_T$ of gluons by three and quarks by two, the result is a ``universal curve'' as shown in the right plot of \Fig{fig:v2qgp}.  The observed scaling is completely accidental, but it led us to consider the possibility of finding similar scaling behavior in a meson / baryon system due to differences in the relaxation rates.  This is discussed in detail in the next section.

\subsection{Two component meson/baryon gas}

\begin{figure}
\begin{center}
 \includegraphics[width=.49\textwidth]{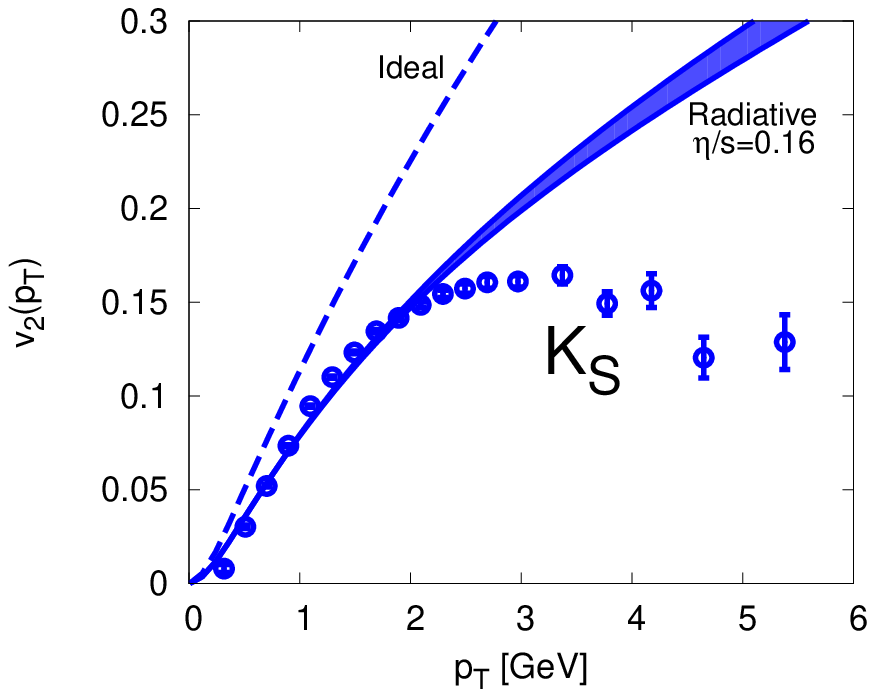}
 \includegraphics[width=.49\textwidth]{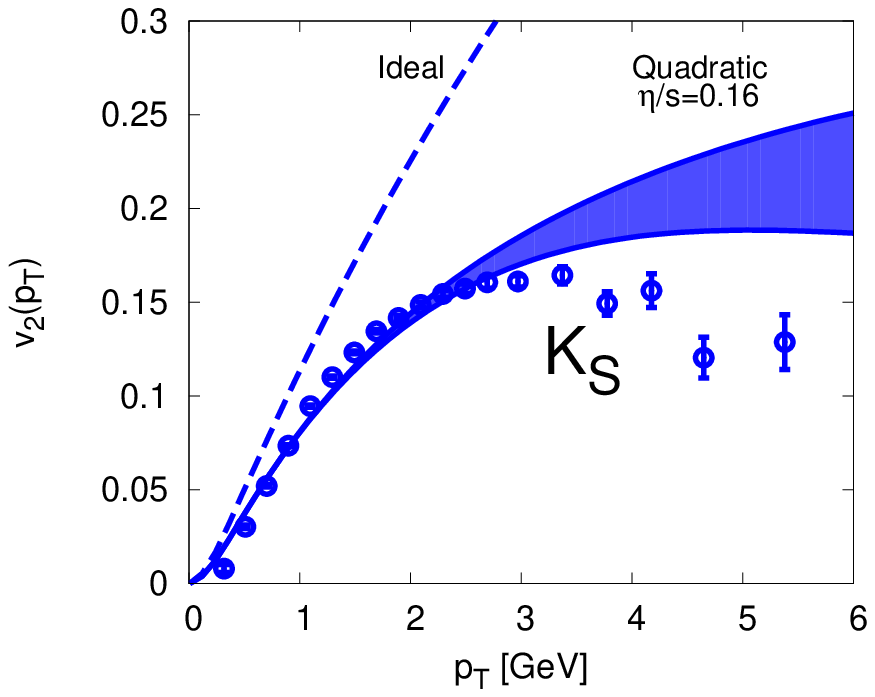}
 \includegraphics[width=.49\textwidth]{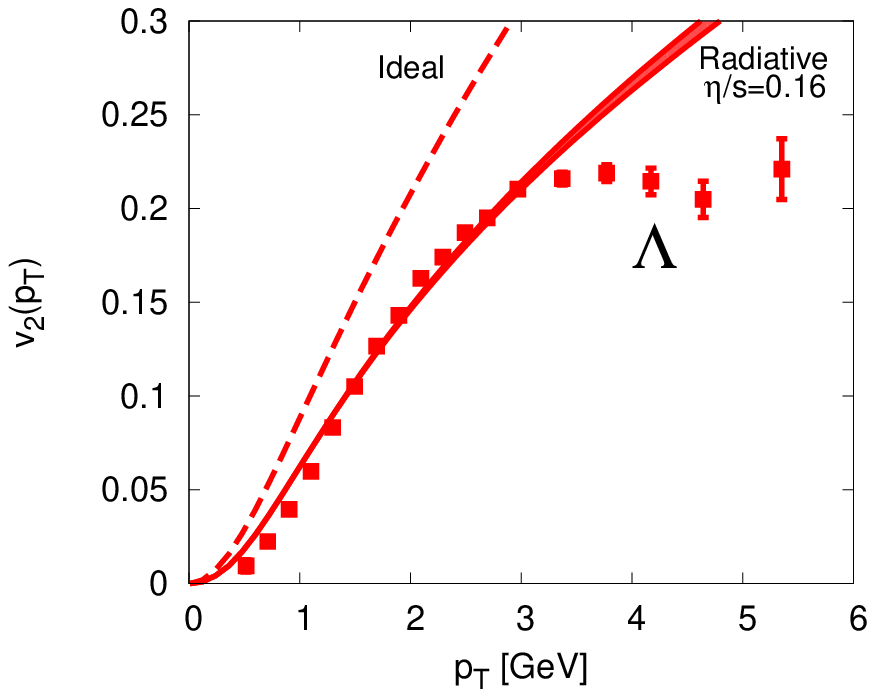}
 \includegraphics[width=.49\textwidth]{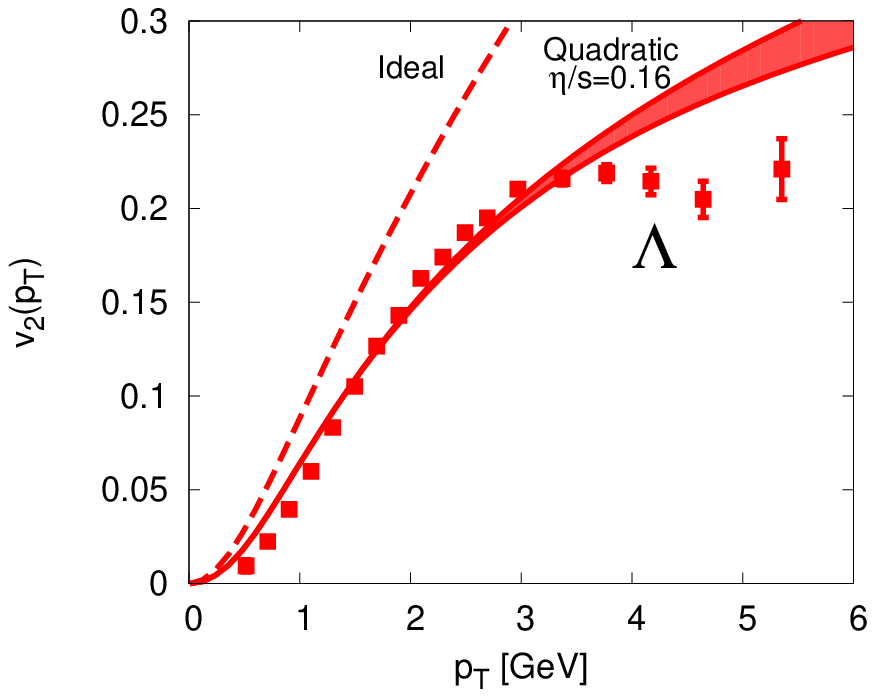}
\end{center}
\caption{Elliptic flow of $K_S$ mesons and $\Lambda$ baryons from viscous hydrodynamics with radiative or quadratic \Ansatze. The run parameters are $\eta/s=0.16$, $T_{\scriptscriptstyle \rm frzout}=150$ MeV and Lattice EoS. Further details are in \app{App:details}. The data is from the STAR collaboration \cite{Abelev:2008ed}.
\label{fig:v2pipradquad}}
\end{figure}

The QCD matter immediately before freezeout is certainly not a weakly
coupled quark-gluon plasma, but it might be described as a hadron
(resonance) gas.  Just as for the quarks and gluons, there is no reason
to think that the mesons and baryons should show the same efficiency in
equilibrating.  But rather than claim a specific model for the partially
equilibrated state of such a system, we will just do some phenomenology
to see how different thermalization rates could affect the observed
species-dependent elliptic flow behavior.  To study the hydrodynamics of
this system we switch from the Ideal gas equation of
state to a lattice motivated \cite{Laine:2006cp}.  Further details of the simulation are presented in \app{App:details}.

We consider a meson / baryon gas whereby mesons and baryons have the
off-equilibrium corrections $f_m$ and $f_b$ respectively,
\beqa
\delta f_{m}(p)&=&-n_p(1+n_p)\chi_{m}(\tilde{p})\hat{p}^i \hat{p}^j\left< \partial_i u_j\right>\,,\nonumber\\
\delta f_{b}(p)&=&-n_p(1-n_p)\chi_{b}(\tilde{p})\hat{p}^i \hat{p}^j\left< \partial_i u_j\right>\,.
\eeqa
We assume both species have the same power-law correction to spectra,
\beqa
\chi_{m}(\tilde{p})&=&C_{m}(\alpha)\tilde{p}^{2-\alpha}\,,\nonumber\\
\chi_{b}(\tilde{p})&=&C_{b}(\alpha)\tilde{p}^{2-\alpha} \,,
\eeqa
but we allow for different coefficients ($C_m / C_b$) which we will choose
in order to give reasonable agreement with data.  For simplicity, we
will consider two different \Ansatze: quadratic ($\alpha=0$) and
radiative ($\alpha=0.5$), and take the following ratios which, as we will
show, fit the data rather well:
\beqa
\frac{C_{m}}{C_{b}}=\left\{ \begin{array}{cc}
1.6 &  \mbox{quadratic,} \\
1.4 & \mbox{radiative.} \end{array}\right.
\eeqa

Finally, the numerical values of the coefficients can be identified with
the shear viscosity through
\beqa
\eta = \frac{1}{15}\sum_{a=\pi,K,...}\nu_a C_{m/b}
   \int\frac{d^3p}{(2\pi)^3 E_a} p^{4-\alpha} n(E_a)\left[1\pm n(E_a)\right],
\eeqa
where the sum extends over all mesons/baryons having $M\le 1.8/2.0$ GeV respectively.
This choice reproduces the lattice parametrization of the equation of
state below $T=160$ MeV.  We find the following values for the
coefficients at our freeze-out temperature of $T=150$ MeV,
\beqa
&&\left.\begin{array}{c}
C_{m}=1.053 \\
C_{b}=0.658\\
\end{array}
\right\}\left(\frac{\eta}{s}\right) \mbox{~~~~~quadratic,} \\
&&\left.\begin{array}{c}
C_{m}=2.661 \\
C_{b}=1.901\\
\end{array}
\right\}\left(\frac{\eta}{s}\right) \mbox{~~~~~radiative.}
\eeqa

Before computing particle spectra we would like to make an aside about
the way elliptic flow is computed.  By definition $v_2(p_T)$ is given by
\beqa
v_2(p_T) \equiv \frac{\int d\phi \cos(2\phi)\:( dN +\delta dN)}
  {\int d\phi \: ( dN +\delta dN )}\,,
\label{eq:v2def}
\eeqa
where $dN$ is  short for $dN/[{dp_T} \,{d\phi}]$ and $\delta  dN$ is
the first viscous correction to this.
In the above expression the viscous correction to the phase space distribution, $\delta dN$, occurs both in the numerator as well as in the normalization from the denominator.  Since we have restricted the viscous correction to be linear in gradients of field quantities we should therefore require that $v_2$ be computed to the same order.  We therefore expand the denominator
\beqa
v_2\approx \frac{\int d\phi \cos(2\phi)\spc dN +\delta dN}{\int d\phi\spc dN} -\frac{\int d\phi \spc \delta dN \int d\phi \cos(2\phi)\spc dN}{(\int d\phi\spc dN)^2},
\label{eq:v2expand}
\eeqa
so the expression retains terms to first order in $\delta f$ only.  In
the following we will show both the expanded and unexpanded expressions
for $v_2$, shading the
region between the two results in order to give an estimate for the
uncertainty in the gradient expansion.  The upper limit of the band
corresponds to \Eq{eq:v2def} while the lower limit is \Eq{eq:v2expand}.
In figures where the uncertainty band is omitted the plotted curve
corresponds to \Eq{eq:v2def}.

Let us now discuss how the different \Ansatze~fare with the experimental
data.  We have chosen $\eta/s=0.16$ in order to give reasonable
agreement with the data in the transverse momentum range $1 \leq p_T
\mbox{ [GeV]} \leq 2$.  The $v_2(p_T)$ spectra for $K_S$ and $\Lambda$
are presented in \Fig{fig:v2pipradquad} using either the radiative
($p_T^{1.5}$) or quadratic ($p_T^2$) \Ansatz.  
  For $p_T \lsim 2$ GeV we find
good agreement between the viscous hydrodynamic results and data.   If
one included hadronic rescattering the low momentum component of the
$\Lambda$ $v_2$ would be pushed out towards higher $p_T$ giving better
agreement with the data.  Above 2-3 GeV large differences between the
radiative and quadratic \Ansatze\ are realized.  We must warn that at
higher $p_T$ one cannot make a direct comparison with data since a
larger fraction of the yield will come from fragmenting partons, which
have not been included.  In addition, the hydrodynamic description
starts to break down at larger $p_T$.  Regardless, one must keep in mind
that for large enough momentum ( {\em i.e.} $p_T \gtrsim 2-3$ GeV) the
two \Ansatze\ used here are clearly discernible and the choice of
\Ansatz\ could in principle lead to differences in the extracted
viscosity.

\begin{figure}
\begin{center}
  \includegraphics[width=.49\textwidth]{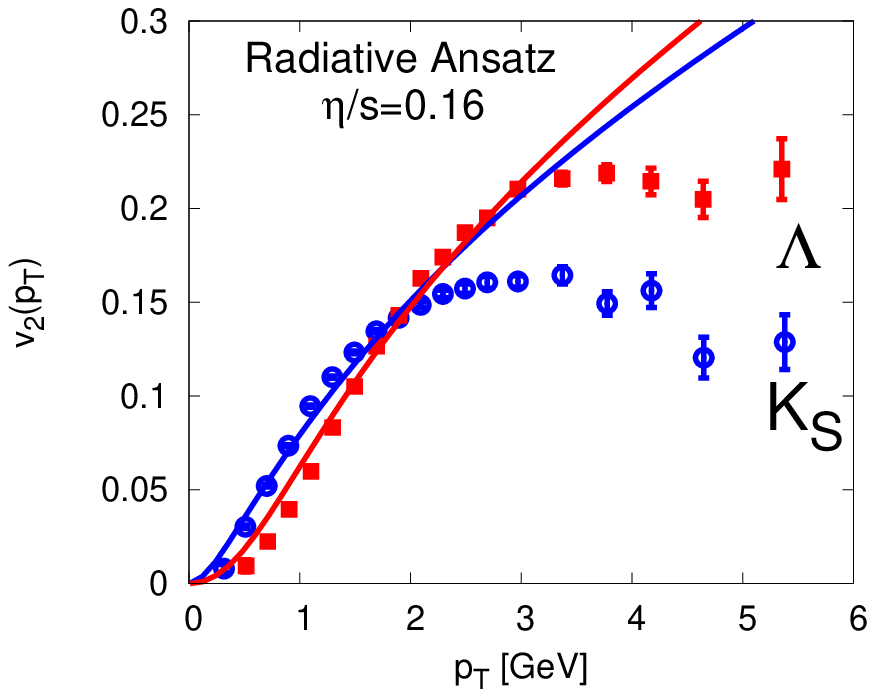}
  \includegraphics[width=.49\textwidth]{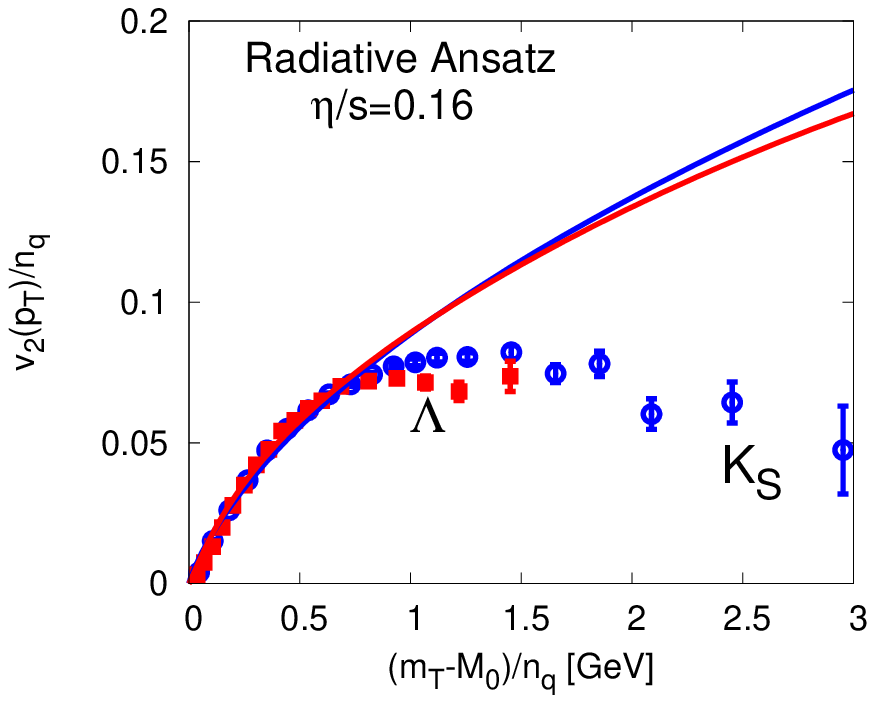}
  \includegraphics[width=.49\textwidth]{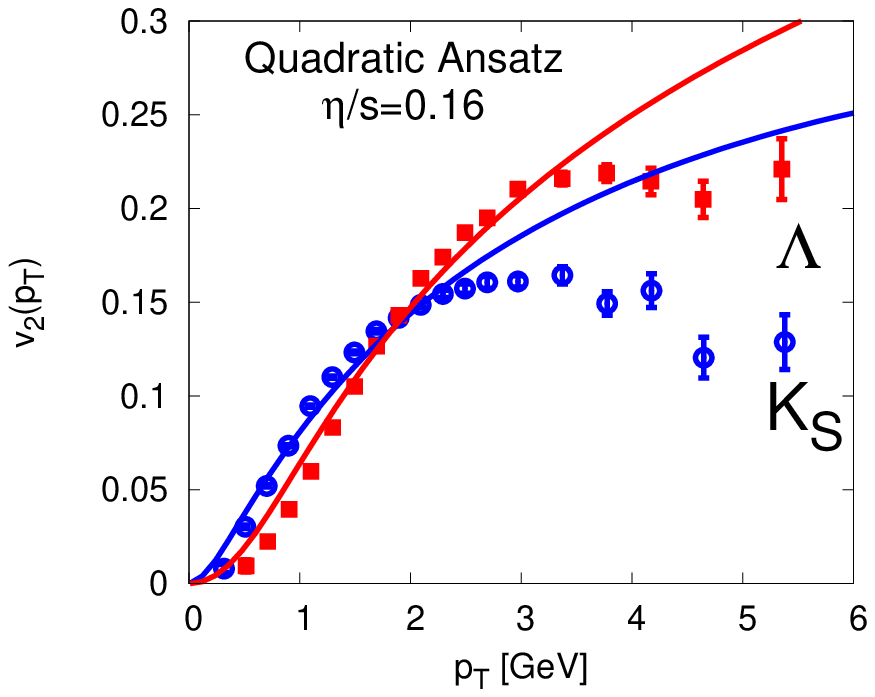}
  \includegraphics[width=.49\textwidth]{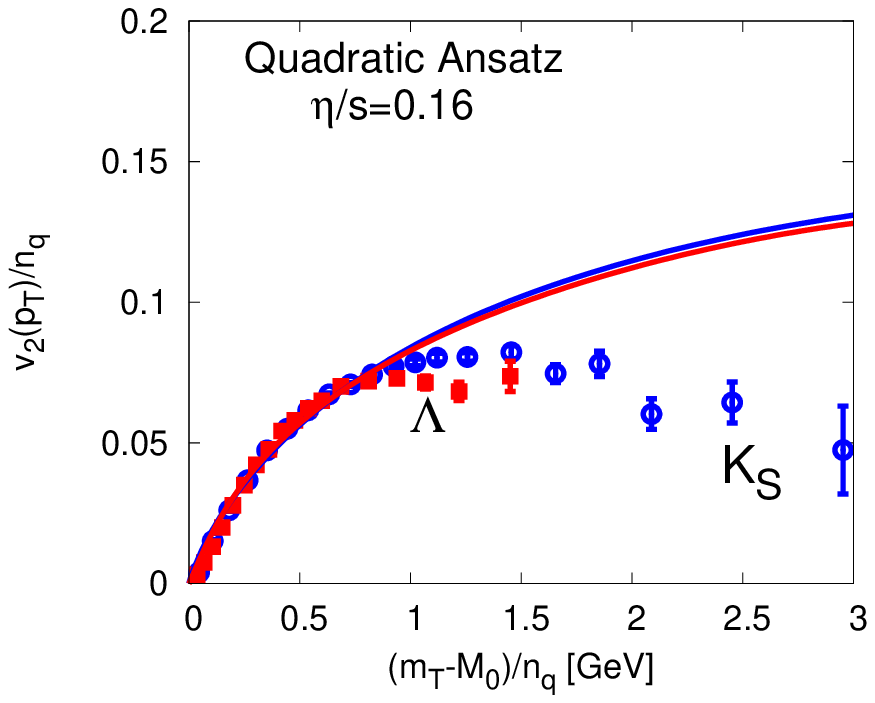}
\end{center}
\caption{Left: $v_2$ of $K_S$ and $\Lambda$. Right: Constituent quark scaling of $v_2(p_T)$. Run parameters can be found in \Fig{fig:v2pipradquad}. The data is from the STAR collaboration \cite{Abelev:2008ed} and is plotted in $(M_T - m)/n$ as suggested by the  PHENIX collaboration \cite{Adare:2006ti}. (Recent PHENIX data presented at Quark Matter \cite{PHENIXv2a,PHENIXv2b} clearly deviate from constituent quark scaling above $(M_T - m)/n \simeq 1$ GeV.)
\label{fig:v2pipscaled}
}
\end{figure}

We would now like to investigate whether we observe a meson / baryon
scaling, similar to the accidental quark / gluon scaling we found from
first principles earlier.  For clarity, we again present the above
results with mesons and baryons on the same figure.  This is shown for
both radiative and quadratic \Ansatze~in \Fig{fig:v2pipscaled}.
The figures show the corresponding results
with both $v_2$ and $p_T$ re-scaled by the number of constituent quarks.
The scaling of the data is the well-known phenomenon of constituent
quark scaling.   We find that viscous hydrodynamics reproduces this
``universal curve'' as well.  This is due to the difference in
relaxation rates between mesons and baryons, which was treated as a free
parameter.   The possible microscopic origin of this ratio is discussed
further in \Sect{summary}.  

We should also point out that this relaxation time scaling is fairly robust to
changes in the equation of state.  While changing the equation of state will
clearly affect the $\eta=0$ behavior, these changes will only have modest
modifications to the viscous correction to the distribution functions. The
qualitative feature that species with smaller relaxation times  have  a
stronger elliptic flow is borne out by \Fig{fig:v2qgp} (a quark gluon plasma
equation of state) and \Fig{fig:v2pipscaled} (a lattice equation of state).
Further study of the equation of state is left to future work.

\section{Summary and Discussion}
\label{summary}

In this work we have presented  a systematic study of the first viscous
correction to the thermal distribution function.  All simulations of viscous
hydrodynamics so far have used the quadratic \Ansatz
\be
 \chi(p) \propto p^2 \, ,
\ee
but this is only an educated guess.

First we studied the form of $\delta f$ (or $\chi(p)$) in a momentum dependent relaxation time approximation and derived a simple formula
\be
 \chi(p) = \tau_R(p) \frac{p}{T} \,.
\ee
Examining this formula we considered two special cases $\tau_R \propto p$
(where the equilibration time is proportional to energy)  and
$\tau_R =\mbox{const}$ (where the equilibration time is independent of energy).
These give rise to quadratic ($\chi(p) \propto p^2$)
and linear ($\chi(p) \propto p$) dependence on momentum,
as is summarized in \Tab{eloss}.  We expect that,
provided QCD is describable in terms of quasi-particles, the first
viscous correction should lie between these cases.
\Fig{fig:v2quadlinear} compares these two extreme
limits for the functional form of the viscous correction.
It is important to emphasize that the two simulations have precisely the
same shear viscosity.  Comparing our results for the elliptic flow in these
two theories, we see that the integrated elliptic flow  $v_2$ is largely
determined  by the shear viscosity, while differential quantities
such as $v_2(p_T)$ at high $p_T$
depend on the equilibration rates at high momentum.
The integrated
elliptic flow is determined  to a large extent by the hydrodynamic
variables $e, u^{\mu}, \pi^{\mu\nu}$.  (An explicit formula
relating $v_2$ to $e,u^{\mu}$ and $\pi^{\mu\nu}$ is
given in \Ref{Teaney:2009qa} which in turn was motivated by earlier observations \cite{KSH,Ollitrault:1992bk,Romatschke:2007mq,Dusling:2007gi}. )
\begin{table}
\begin{tabularx}{\textwidth}{
 c  @{\hspace{0.1in}} | @{\hspace{0.1in}} X @{\hspace{0.1in}} | @{\hspace{0.1in}} l}
 Model           &    Physics  &    Formula \\ \hline
 Relaxation time, $\tau_R \propto p$ &  Relaxation time grows with particle momentum.     & $\chi(p)  \propto p^2$ \\
Relaxation time , $\tau_R = \mbox{const}$ &  Relaxation time independent of momentum.  & $\chi(p) \propto  p $\\ \hline  \hline
 Scalar theory &  Randomizing collisions which happen rarely & $\chi(p) \propto p^2$ \\ \hline \hline
 QCD Soft Scatt.   & Soft $q \sim gT$ collisions lead to a random walk of
hard particles. &  $\chi(p) \propto  p^2$  \\
 QCD Hard Scatt.
& Hard  $q \sim \sqrt{pT} $ collisions lead to a random walk of
hard particles.
&  $\chi(p) \propto \frac{p^2}{\log(p/T) }$ \\
 QCD Rad. E-loss  & Radiative energy controls the
approach to equilibrium. In the LPM regime $\hat{q}$
controls the radiation rate.  &   $\chi(p)  \propto  \frac{p^{3/2}}{\alpha_s \sqrt{\hat q}}$ \\ \hline
\end{tabularx}
\caption{ \label{eloss}
Summary of the functional dependence of the departure from equilibrium
on the theory and approximation considered. }
\end{table}

The quadratic ansatz is valid only for fairly specialized theories.
For instance, examining \Tab{eloss} we see that scalar theories follow
this \Ansatz.  The reason is that the cross-section falls as $1/s$, so
higher-energy particles see a more transparent medium and equilibrate more
slowly.

For different reasons
the quadratic \Ansatz\ is also valid in a soft scattering approximation
to high temperature QCD (see Row 4 of \Tab{eloss}).  In this limit, which
treats  $\log(T/m_D)$  as an expansion parameter, soft $g T$ collisions
lead to the momentum diffusion and drag of hard gluons. 
If the momentum diffusion is independent of particle energy and the drag
is constant, we get the quadratic \Ansatz.  If the momentum diffusion
increases logarithmically with particle energy, we find a logarithmic
correction to this \Ansatz\ (see Row 5 of \Tab{eloss}). A formula which 
summarizes the asymptotic form of both of these cases is 
\be
   \chi(p) = \frac{p^2}{2 T\llangle dE/dt \rrangle_p} \, ,
\ee
where $\llangle dE/dt\rrangle_p$ 
is the rate is energy loss  of a particle 
with momentum $p$ (see \Eq{bthoma_soft} and \Eq{bthoma} for 
explicit formulas in certain limits).

However, the effect of bremsstrahlung completely changes this picture.
A naive (Bethe-Heitler) treatment of radiative energy loss would
lead to a relaxation rate independent of momentum, but including the LPM
effect, the viscous correction behaves asymptotically as
\be
\label{qhat_chi_conc}
  \chi(p)  = 0.7 \frac{ p^{3/2} }{\alpha_s \sqrt{\hat{q}}} \, .
\ee
This formula is summarized in Row 6 of \Tab{eloss} and provides a concrete 
connection between  viscous corrections and radiative energy loss 
which is further explored in \Fig{fig:eloss} and surrounding text. 

From a phenomenological perspective, the
LPM effect is not entirely dominant and collisions are important in
the relevant momentum range. A phenomenological fit
to numerical results for the first viscous correction,
including  both collisions and collinear radiation without making the
strict LPM approximation,
shows that the first viscous correction is reasonably well described by the
following phenomenological form:
\be
  \chi(p) \simeq C \tilde p^{1.38} \, .
\ee
\Fig{fig:v2pureglue} compares this functional form to the linear and
quadratic \Ansatze\
motivated by the relaxation time approximation. We see that the general
expectation from high temperature QCD is that in the relevant momentum
range the first viscous correction is slightly closer to the  linear
rather than the quadratic ansatz.

We next studied two
component plasma starting with a two component plasma of quarks and
gluons.  Since the relaxation rates of the quarks and gluons
are not the same the two components do not have the same distribution
function. At high momentum an analysis of collinear splittings
$g\rightarrow gg$, $g\rightarrow q\bar{q}$, $q\rightarrow gq$ shows
that both the quark and gluon distribution behave as $p^{3/2}$.
However the ratio of the quark and gluon viscous corrections
approaches a constant
\be
\label{quark_glue_17}
  \frac{\chi_q}{\chi_g} \approx 1.70  \,.
\ee
The constant is determined by the ratio of Casimirs $C_A/C_F=9/4$
and the dynamics of the QCD splitting functions. It also depends weakly
on the number of quark flavors  and we have quoted the two flavor case.

Motivated by this example, we have postulated that the baryon and meson
components of the medium have different equilibration rates. Indeed,
there is no reason to expect that these species would equilibrate at the
same rate.  Then we fitted (by eye)
the ratio of relaxation rates to reproduce the baryon and meson
elliptic flows. If the ratio of relaxation times is
\be
  \frac{\chi_m}{\chi_b} \simeq 1.5 \,,
\ee
meaning that baryons relax to equilibrium 1.5 times faster than mesons,
then the resulting viscous hydrodynamic calculation effortlessly reproduces
the universal ``constituent quark scaling'' curve. Physically what is
happening is that in ideal hydrodynamics the baryons and mesons have
approximately the same elliptic
flow which is approximately described by a linear rise in $m_T$.  The viscous correction then dictates that the baryons will follow this ideal trend 1.5 times farther than the mesons. 
Although it is not obvious from the data shown 
in 
\Fig{fig:v2pipscaled}, the data do not show scaling above 
$(m_T - M_o)/n_q\simeq 1\,{\rm GeV}$, {\it i.e.}  the last Lambda
point is a fluctuation upward (This is seen quite clearly in recent 
PHENIX data\cite{PHENIXv2a,PHENIXv2b}.) It is interesting that the data also
deviate from hydrodynamic predictions above this point.

It is tempting to speculate as to the microscopic origin of the factor of
$1.5$.  The baryons and mesons in the $2-3\,{\rm GeV}$ region are produced in
the complex transition region where the energy density decreases from
$1.2\,{\rm GeV}/{\rm fm^3}$ to $0.5\,{\rm GeV}/{\rm fm^3}$.  In this range, the
temperature decreases by only $\Delta T \simeq 20\,{\rm MeV}$. However, the
hydrodynamic simulations evolve this complicated region  for a significant
period of time, $\tau \simeq 4\,{\rm fm} \leftrightarrow 6.5\,{\rm fm}$,
and the hadronic currents are built up over this time period.   The
interactions are probably quite inelastic and are not easily classified as
hadronic or partonic in nature.  The additive quark model was used to describe
high energy total cross sections which are similarly inelastic \cite{AQM}. It predicts
the ratio of high energy nucleon-nucleon to pion-nucleon (as well as pion-nucleon to pion-pion) cross sections to be $3/2$ in reasonable agreement with the experimental ratio.
Perhaps
similar physics is responsible for the different baryon and meson elliptic
flows.  In fact, the 
splitting of the baryonic and mesonic elliptic flows
was predicted at least qualitatively by UrQMD which implements the additive quark model \cite{Bleicher:2000sx}. 
On the other hand, 
 the factor of $1.5$ in the relative relaxation times  could be
simply a combination of dynamical and group theoretical factors of accidental
significance.  

In summary, a species dependent relaxation time 
provides a coherent and physically transparent explanation for the complicated
trends observed in the elliptic flow data  measured at RHIC.

\section*{Acknowledgments}
KD is supported by the US-DOE grant DE-AC02-98CH10886. DT would
like to thank Paul Sorenson, Raimond Snellings, and Jiangyong Jia 
for informative discussions.
DT is supported in part by an OJI grant from the US Department of Energy DE-FG-02-08ER4154 and the Sloan Foundation.  
GM would like to thank Paul Romatschke for useful conversations, the
physics department at the Universidad Autonoma Madrid for hospitality
while this work was completed, and the Alexander von Humboldt Foundation
for its support through a F.\ B.\ Bessel prize.  GM's work was supported in part by the Natural Sciences and Engineering Research Council of Canada.

\appendix

\section{Details of hydrodynamic description}
\label{App:details}

The initial condition of the hydrodynamic evolution is set by a Glauber
model and the energy density is proportional to the number of binary
collisions.  More specifically we take
\beqa
\epsilon(\tau_0=1\mbox{ fm},x,y)=
    E_{BC}\times\frac{n_{coll}(b,x,y)}{\sigma_{NN}}
\eeqa
where $E_{BC}=22.735$ is the energy per binary collision and
$\sigma_{NN}=40$ mb is the inelastic nucleon-nucleon cross section.

In this work we will use the following evolution equation for $\pi^{ij}$,
\begin{eqnarray}
\dot{\pi}^{ij}&=&-\frac{1}{\tau_{\pi}}(\pi^{ij}-\eta\sigma^{ij})
  -2\pi^{ij}\partial_k u^k + \pi_{k(i}\omega_{j)}^k
   +\frac{1}{\eta}\pi_{k\langle i}\pi_{j\rangle}^k,
\label{eq:evol2}
\end{eqnarray}
which is identical to the stress tensor used in \cite{Dusling:2007gi}.
Other possibilities are also possible \cite{Dusling:2009zz} which will
not change the results of this work on a qualitative level.  In the
above expression $\omega_{ij}\equiv\partial_j u_i-\partial_i u_j$ is the
vorticity and $\tau_\pi=3\eta/(4p)$.
There is one technical detail that warrants discussion.  At large transverse distances the viscous pressure tends to become larger than the ideal pressure and the equations become unstable.  It is therefore necessary to cutoff our auxiliary tensor when it becomes large.  More precisely we take 
\beqa
\pi^{ij}\to\frac{\pi^{ij}}{1+\kappa\mbox{Tr}\pi^2},
\eeqa
where $\mbox{Tr}\pi^2=\sqrt{\pi_{11}^2+\pi_{22}^2+\pi^2_{33}}$ and $\kappa\approx 0.1/(\alpha p)$.

In the first part of this paper we consider an ideal gas equation of state, $p=1/3\epsilon$.
For a two flavor QGP the ideal Stefan Boltzmann gas gives $\epsilon = 12.71 T^4$, which
roughly corresponds to the $\epsilon/T^4$ relation found on the lattice.  (For the highest temperatures
in the simulation it is above this value and for the lowest temperatures in the simulation it is this value).  We have decided to use the same $\epsilon/T^4$ ratio for both the gluon gas and  
quark + glue simulations in order to get the fairest possible phenomenological estimate for the
size of the viscous corrections in a realistic heavy ion event.

For simulations using the ideal gas EoS the freeze-out contour is taken at constant $\epsilon_{\scriptscriptstyle \rm frzout}=0.6$ GeV/fm$^3$ corresponding to a temperature of 140 MeV.  The default impact parameter is 7.6 fm and the shear viscosity to entropy ratio is $\eta/s=0.08$.

In the second part of this paper where we compute spectra of a meson/baryon gas we have used a lattice motivated equation of state \cite{Laine:2006cp}.  In this case the freeze-out surface is set by $\epsilon_{\scriptscriptstyle \rm frzout}=0.24$ GeV/fm$^3$ corresponding to a temperature of 150 MeV.  We have used a default impact parameter of 6.8 fm corresponding to a centrality class of 10-40\% and a shear viscosity to entropy ratio of $\eta/s=0.16$.

\section{Collision Integrals}
\label{collision}

In this section we will give the details leading to \Eq{scalar_res}
for a scalar theory and \Eq{chi_coll} for pure glue.

\subsection{Scalar theory}

Our starting point is \Eq{22cofdf}. Substituting
the form specified in \Eq{eq:phiquad} into this equation yields in
a Boltzmann approximation
\beqa
\frac{\llangle \partial_i u_j \rrangle  p_i p_j}{T} \frac{e^{-p/T}}{p} & = &
\frac{C \llangle \partial^{\mu} u^{\nu} \rrangle }{T^3}
  \frac{e^{-p/T}}{2p} \int \frac{d^3 k}{(2\pi)^3 2k}e^{-k/T}
\,  \frac{\lambda^2}{2}
 \int \frac{d^4 \P' d^4 \K'}{(2\pi)^2}\delta(\P'{}^2) \delta(\K'{}^2)
\nonumber \\ && \times \,
\delta^4( (\P{+}\K) - \P'-\K')
\, \left( \P_\mu \P_\nu + \K_\mu \K_\nu
    - \P'_{\mu} \P'_{\nu} - \K'_{\mu} \K'_{\nu} \right) \,,
\eeqa
where $\llangle \partial^{\mu} u^{\nu} \rrangle$ is the Lorentz invariant
extension\footnote{
Specifically, defining the projector onto the local rest frame
$\Delta^{\mu\nu}=g^{\mu\nu} + u^{\mu}u^{\nu}$,  we
have
\[
\llangle \partial^{\mu} u^{\nu} \rrangle = \frac{1}{2}\Delta^{\mu\rho}
\Delta^{\nu\sigma} \left(\partial_\rho u_\sigma + \partial_\sigma u_\rho -
\frac{2}{3}\Delta_{\rho\sigma} \partial_\beta u^{\beta} \right) \, .
\]
In the local rest frame implicit here, we have $\llangle \partial^{0} u^{\mu} \rrangle=0$ and $\llangle \partial^\mu u^{\nu}\rrangle =\llangle \partial^{i}u^{j} \rrangle $ for $\mu,\nu=1..3$.
}
of $\llangle \partial^i u^j \rrangle$.
The integrals over $P'$ and $K'$ are Lorentz covariant and can be performed by standard
tricks; the $(\P_\mu \P_\nu+ \K_\mu \K_\nu)$ term can be factored out, leading to
$\int_{\P',\K'}=1/8\pi$, while the integral over
$\P'_\mu \P'_\nu + \K'_\mu \K'_\nu$ must return a rank-2 tensor depending
only on $(\P{+}\K)_\mu$.  There are only two such tensors, and contraction
with $g^{\mu\nu}$ and $(\P{+}\K)^\mu (\P{+}\K)^\nu$ establishes
that
\begin{align}
\int\! \frac{d^4 \P' d^4 \K'}{(2\pi)^2}  \delta(\P'{}^2) \delta(\K'{}^2)
\delta^4( (\P{+}\K) &- \P'-\K')
\left( \P'_{\mu} \P'_{\nu} {+} \K'_{\mu} \K'_{\nu} \right) \nonumber \\
&= \frac{1}{48\pi} \left( 4 (\P{+}\K)_\mu (\P{+}\K)_\nu
   - (\P{+}\K)^2 g_{\mu\nu} \right) \, .
\end{align}
The integral equation becomes
\beqa
\frac{\llangle \partial_i u_j \rrangle p_i p_j}{T} & = &
\frac{C \llangle \partial_i u_j \rrangle \lambda^2}{32 \pi T^3}
 \int \frac{d^3 k}{(2\pi)^3 2k}e^{-k/T}
\left( p_i p_j + k_i k_j - \frac{2}{3} (p+k)_i (p+k)_j \right) \,,
\eeqa
where we used that $\llangle \partial^\mu u^\nu \rrangle$ is traceless,
$\llangle \partial^{\mu} u^{\nu} \rrangle g_{\mu\nu}=0$.  Performing the $k$ angular integration in the plasma
frame, the $p_i k_j$ terms integrate to zero; so does the $k_i k_j$
term, because $\llangle \partial_i u_j \rrangle $ is traceless.  Performing the trivial radial
integration, we find \Eq{scalar_res}.

\subsection{Pure glue}

Our goal here is to derive \Eq{chi_coll} and \Eq{bthoma}.
Our starting point is the collision integral \Eq{eq:Gamma}  with
matrix elements given by \Eq{megg}
\be
 \frac{p^i p^j}{T E_p} \llangle \partial_i u_j \rrangle
 = \int_{\p'\kk\kk'} n_p n_k (1 + n_{k'}) (1 + n_{p'} )  \Gamma_{pk\rightarrow p'k'}
\left[\chi(\p) + \chi(\kk) - \chi(\p') - \chi(\kk') \right]\,.
\ee
Using the definition, $\chi(\p) = \chi(p) \left(\hat{p}^i \hat{p}^j  - \delta^{ij}/3 \right) \llangle \partial_i u_j \rrangle$, one can pull out
the common factor, $\llangle \partial_i u_j \rrangle$. The remaining
integral on the right hand side (called $I^{ij}$) must have the form
$I^{ij} = I(p) \left(\hat{p}^i\hat{p}^j - \delta^{ij}/3\right)$ since this is
the only symmetric traceless tensor which can be constructed out of
$\p$ and $\delta^{ij}$.  Straightforward analysis  then shows that
\beqa
 \frac{p n_p(1 + n_p)}{T} &=& \int_{\p'\kk\kk'} n_p n_k(1+ n_{k'}) (1 + n_{p'})
\,  \Gamma_{pk\rightarrow p'k'}
\\
& & \quad  \times \;  \left[ \chi(p) + \chi(k) P_2(\cos\theta_{\p\kk}) -  \chi(p') P_2 (\cos\theta_{\p\p\pr})  - \chi(k') P_2(\cos\theta_{\p\kk\pr})  \right]  \, , \nonumber
\label{eq:coll3}
\eeqa
where for instance
\be
P_2(\cos\theta_{\kk\p}) = \frac{3}{2} \left(\hat{p}^i \hat{p}^j - \frac{1}{3} \delta^{ij} \right) \left(\hat{k}^i \hat{k}^j - \frac{1}{3} \delta^{ij} \right) \, ,
\ee
is the second Legendre polynomial.

\begin{figure}
\includegraphics[width=0.55\textwidth]{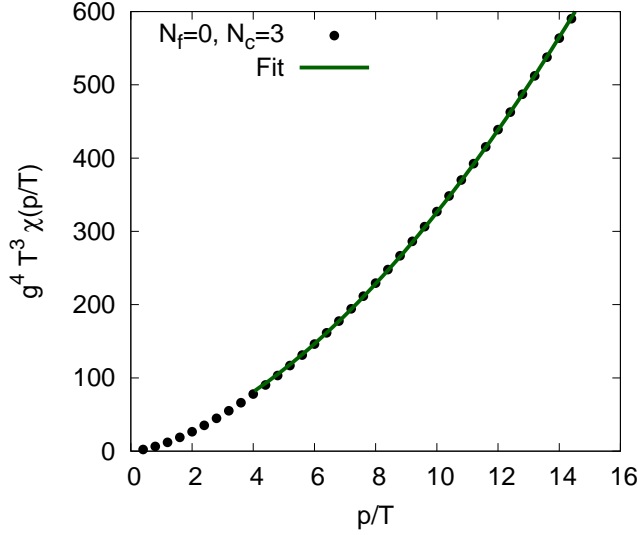}
\caption{Off-equilibrium correction for the case of collisional energy loss.
The points are from the numerical solution of the linearized Boltzmann equation and the
curve is the asymptotic form, \Eq{chicoll3}. Specifically the curve is a one
parameter fit to the $N_c=3$ form, $(24\pi/9) \; \tilde p^2/\log(\tilde p/C)$, with fit parameter
$C^{-1}=1.3$. }
\label{fig:chicoll}
\end{figure}

We will evaluate this integral in a leading $\log(p/T)$ approximation.
Asymptotically, the momenta $p$ and
$p'$
are large,  while $k$ and $k'$ are of order the temperature\footnote{We
will discuss the region of phase space where $t=-(P'-P)^2$ is small.
Since the particles are identical, there is also an equal contribution
where $u=-(K'-P)^2$ is small, {\it i.e.} when $p$ and $k'$ are large and $p'$ and $k$ are of order
$T$.  Our original
definition of the transition rate includes a $1/2$ symmetry factor for
the identical particle final state. To ease the discussion in this
section, we will simply drop the symmetry factor and neglect $u$-channel
contribution. }.
In this limit we can make the
Boltzmann approximation, $\left(1+n_{p'}\right) \rightarrow 1$,
and can treat $p'$ as  close to $p$. Specifically
we take
\be
 \cos\theta_{pp'} = 1 +  \frac{t}{2 pp'} \simeq 1  \,,
\ee
and then write
\be
 \chi(p) - \chi(p') \simeq -\frac{\partial \chi}{\partial p}\, \omega \, .
\ee
We also note that $k$ and $k'$ are close to $T$
and therefore $\chi(k)$ and $\chi(k')$ are small.
Then we can write \Eq{eq:coll3} as
\be
 \frac{p}{T}  \simeq   \frac{\partial \chi}{\partial p} \llangle \frac{dE}{dt} \rrangle_p \, ,
\ee
where the average energy loss rate for a particle with momentum $p$ is
\be
 \llangle \frac{dE}{dt} \rrangle_p  = - \int_{\p'\kk\kk'} \Gamma_{\p\kk \rightarrow \p'\kk'} n_k (1 + n_{k'}) \omega \, ,
\ee
{\it i.e.}
the energy loss is the transition rate weighted with the energy transfer.
The energy loss to leading $\log(p/T)$
has been determined by Bjorken \cite{Bjorken:1982tu} and Braaten and Thoma \cite{Braaten:1991jj} and  reads
\be
\label{lleloss}
 \llangle \frac{dE}{dt} \rrangle_p \simeq \frac{g^4 C_A^2 T^2}{48 \pi}\log\left(\frac{p}{T}\right)\, .
\ee
One can verify that when terms suppressed by $\log(p/T)$ are dropped
we have
\be
\label{chicoll3}
   \chi(p) = \int_{\sim T}^{\sim p} \frac{p'}{T\llangle {dE}/{dt}\rrangle_{p'}}\, {\dd p'} \,   \simeq
\frac{1}{T\llangle {dE}/{dt}\rrangle_p } \frac{1}{2} p^2 \, .
\ee
\Fig{fig:chicoll} shows a fit based on \Eq{chicoll3} which does
a reasonable job in reproducing our $N_c=3$ numerical results at high momentum.

\begin{figure}
\includegraphics[width=0.48\textwidth]{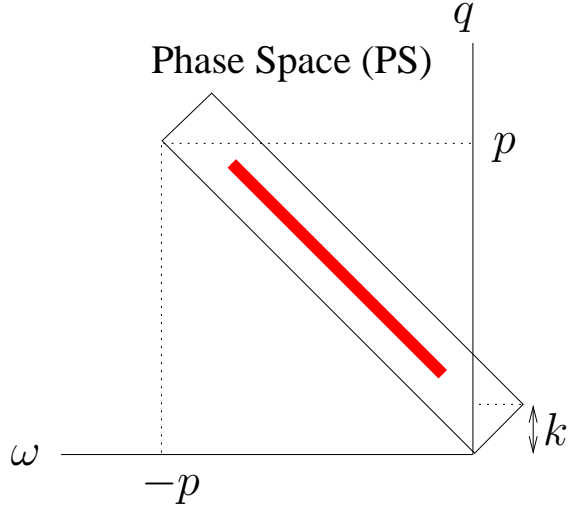}
\caption{The available phase space for the collision integrals in
  \protect\Eq{phasespacei}. The band shows the dominant region of the
  integration in a leading $\log(p/T)$ approximation.}
\label{fig:phasespace}
\end{figure}

For completeness we will rederive \Eq{lleloss}.
In order to evaluate the phase space integrals over $\Gamma_{\p\kk\rightarrow \p\pr\kk\pr}$
we use the ``t-channel parametrization"
of \Ref{Arnold:2003zc}.  Following the logic that leads from (A.14) to (A.21) of
this work, we write the phase space as
\beqa
\label{phasespacei}
\int_{\p'\kk\kk'} \Gamma_{\p\kk\rightarrow \p'\kk'}
 = \frac{1}{(2\pi)^4  16p^2} \int_0^{\infty} dk \int_0^{2\pi} d\phi \int_{PS} dq \int_{PS}  d\omega  \left|\mathcal M\right|^2 \, ,
\eeqa
where the momentum transfer is  $\q = \p' - \p = \kk - \kk'$  and the
energy transfer is $\omega = p' - p = k - k'$. The vector
$\q$ is taken along the $z$ axis and the vector $\p$ lies
in the $z-x$ plane. The angle $\phi$ is the azimuthal
angle of $\kk$ with respect to the $z,x$ plane. The energy transfer
and momentum transfer are restricted to the available phase space
\beqa
 0 &<& \frac{q+\omega}{2}   < k \, , \\
 0 &<& \frac{q-\omega}{2}   < p \, ,
\eeqa
which is also exhibited in \Fig{fig:phasespace}.

Using the definitions of the kinematic variables, the Mandelstam invariants
are
\beqa
 t &=& -(P'-P)^2 = q^2 - \omega^2 \, , \\
 s &=& -(P+K)^2 = \frac{-t}{2q^2}\left[ (p +p')(k+k') + q^2 - \cos\phi \sqrt{(4pp' + t)(4kk' + t) } \right] \, , \\
 u &=& -t - s  \, .
\eeqa
To evaluate the collision integral, we are to substitute these
expressions for the Mandelstam invariants into the matrix elements
and perform the integrals over the phase space. Close inspection
of the result of this procedure shows how $\log(p/T)$ comes about. First, the logarithm comes from integrating over the phase space region
where $\omega\simeq -q$ and
$T \ll q \ll p$  as
shown by the band shown in \Fig{fig:phasespace}.
Since the $\omega$ integral is over  the interval, $-q < \omega < -q + 2k$,
 the phase-space integral
is approximately
\be
 \int_{PS} dq \int_{PS} d\omega \simeq  2k \int_{\sim T}^{\sim p} dq\,,
\ee
and we may neglect the stimulation factor, $(1 + n_{k'} ) \simeq 1$.
Second, only the
highest powers of $\omega$ and $q$ contribute  to the ultraviolet logarithm.
The $\phi$ integrated matrix elements with these restrictions is
\beqa
 \int_0^{2\pi} d\phi\,  \frac{-u s}{t^2}   
&\simeq &  \frac{2\pi p^2}{q^2}\,.
\eeqa
Then the total total transition rate
is
\be
\llangle \frac{dE}{dt} \rrangle_p
= - \frac{1}{(2\pi)^4 16 p^2}  \left[8 g^4 C_A^2 \right] \int_0^{\infty}  dk\, 2 k\,  n_k \,  \int_{\sim T}^{\sim p} dq  \frac{2\pi p^2}{q^2} (-q) \, .
\ee
Performing the integral over $k$, we arrive at the result quoted in \Eq{lleloss}.

\section{Viscous distribution function and $\hat{q}$}
\label{app:qhat}

In this appendix we derive the form of the viscous distribution function
for asymptotically large momenta.  In doing this we will relate the high
$p_T$ tail of the distribution function with the energy loss parameter
$\hat{q}$.

The starting point is the Boltzmann equation containing near collinear
splitting processes.  We neglect $2 \leftrightarrow 2$ processes as
these will be sub-leading at large momenta.  Therefore
\beqa
\frac{p^\mu}{E_p} \partial_\mu f_a({\bf x},{\bf p})=-\mathcal{C}^{1\to 2}_a[f] \, ,
\eeqa
where
\be
\mathcal{C}^{1\to 2}_a=\frac{(2\pi)^3}{2\vert{\bf p}\vert^2 \nu_a}
  \int_0^\infty dp\pr dk\pr \delta(\vert{\bf p}\vert-p\pr-k\pr)
  \gamma({\bf p};p\pr,k\pr)
  \left[f_p(1\pm f_{p\pr})(1\pm f_{k\pr})-f_{p\pr}f_{k\pr}(1\pm f_p)\right]\,.
\ee
In the above expression $f_a$ is the distribution function of species
$a$.  The degeneracy factor, $\nu_a$ is 16 for gluons and $6$ for
quarks.

Now linearize the collision integral
\beqa
\mathcal{C}^{1\to 2}=\frac{(2\pi)^3}{2p^2 \nu_g}\int_0^\infty dp\pr dk\pr \delta(p-p\pr-k\pr)\gamma(p;p\pr,k\pr)n_p(1\pm n_{p\pr})(1\pm n_{k\pr})\left[\chi_p-\chi_{p\pr}-\chi_{k\pr}\right]\,.
\eeqa
Doing the integral over $k\pr$ and expanding out the LHS in the typical
way we get
\be
\beta n_p(1\pm n_p)\frac{p^2}{E_p}=-\frac{(2\pi)^3}{2p v_a}\int_0^\infty dx
\gamma(p;xp,(1-x)p)n_p(1\pm n_{xp})(1\pm
n_{(1-x)p})\left[\chi_p-\chi_{xp}-\chi_{(1-x)p}\right]\,.
\ee
Then note at very high momentum
\beqa
\frac{(1\pm n_{xp})(1\pm n_{(1-x)p})}{1\pm n_p}\to \Theta(1-x)\,.
\eeqa

Let us now consider a two component plasma of quarks and gluons.
Using the \Ansatz\ $\chi_{q,g}(p)=C_{q,g}p^{2-\alpha}$ we are left with
\beqa
\label{eq:C6}
\frac{p^2 \nu_g}{(2\pi)^3}&=&\frac{1}{2} p^{2-\alpha}
   \int_0^1 dx \gamma^g_{gg}(p;xp,(1-x)p)
  \left[C_g-C_g x^{2-\alpha}-C_g (1-x)^{2-\alpha}\right]
 \nonumber\\
&& +p^{2-\alpha} \int_0^1 dx \gamma^g_{q\overline{q}}(p;xp,(1-x)p)
 \left[C_g-C_q x^{2-\alpha}-C_q (1-x)^{2-\alpha}\right],
\nonumber\\
\frac{p^2 \nf \nu_q}{(2\pi)^3}&=& p^{2-\alpha} \int_0^1 dx
  \gamma^q_{gq}(p;xp,(1-x)p)
  \left[C_q-C_g x^{2-\alpha}-C_q (1-x)^{2-\alpha}\right].
\eeqa

The splitting functions at leading log order are
\beqa
\gamma^g_{gg}(p;xp,(1{-}x)p)&=&\frac{\sqrt{6}\alpha_s C_A d_A}{(2\pi)^4}
  \sqrt{p\hat{q}}\frac{\sqrt{C_A+C_Ax^2+C_A(1{-}x)^2}
 \left[1+x^4+(1{-}x)^4\right]}{\left[x(1{-}x)\right]^{3/2}},
\\
\gamma^g_{q\overline{q}}(p;xp,(1{-}x)p)&=&
  \frac{\sqrt{6}\alpha_s C_F d_F \nf}{(2\pi)^4}
  \sqrt{p\hat{q}}\frac{\sqrt{(2C_F-C_A)+C_Ax^2+C_A(1{-}x)^2}
  \left[x^2+(1{-}x)^2\right]}{\left[x(1{-}x)\right]^{1/2}},
\nonumber\\
\gamma^q_{gq}(p;xp,(1{-}x)p)&=&
  \frac{\sqrt{6}\alpha_s C_F d_F \nf}{(2\pi)^4}
  \sqrt{p\hat{q}}\frac{\sqrt{C_A+(2C_F-C_A)x^2+C_A(1{-}x)^2}
  \left[1+(1{-}x)^2\right]}{x\left[x(1{-}x)\right]^{1/2}}. \nonumber
\eeqa
The $p^{1/2}$ behavior here together with the $p^{2-\alpha}$ behavior
explicitly on the RHS of \Eq{eq:C6} must cancel the $p^2$ behavior on
the LHS of \Eq{eq:C6}.  This fixes $2 = 2-\alpha + 1/2$ or $\alpha=1/2$,
so $\chi(\tilde p)\propto \tilde p^{3/2}$.  This proves the claim in the
main text that the asymptotic behavior should be $\alpha = 1/2$.

%

For a gluon gas we find
\beqa
C_g^{-1}=\frac{C_A^{3/2} \alpha_s \sqrt{3\hat{q}}}{2\pi T}
  \int_0^1\frac{\left[1-x(1-x)\right]^{5/2}}
  {\left[x(1-x)\right]^{3/2}}
  \times\left[1-x^{3/2}-(1-x)^{3/2}\right]dx \,,
\eeqa
and
\beqa
\chi_g=\frac{0.704778}{\alpha_s T \sqrt{\hat{q}}}p^{3/2}\,.
\eeqa
For a two-flavor quark-gluon gas we find
\beqa
\chi_g=\frac{0.759158}{\alpha_s T \sqrt{\hat{q}}}p^{3/2},\nonumber\\
\chi_q=\frac{1.257913}{\alpha_s T \sqrt{\hat{q}}}p^{3/2}.
\eeqa
The ratio is
\beqa
\frac{\chi_q}{\chi_g}=1.657 \,,
\eeqa
not too different from the ratio $1.7$ we found by fitting.
This ratio depends on $\nf$.  For 1 flavor it is 1.702, for 3 flavors it
is 1.618, and in the limit of infinite flavors it approaches 1.128.
This diminishing ratio occurs because, at larger $\nf$, more and more
splitting processes are $\GQQ$ and $\QQG$, which equilibrate the numbers
of quarks and gluons towards each other.

\section{Two component system}
\label{app:twocomp}

In this appendix we derive relationships between the off-equilibrium distribution function, $\chi$ and the shear viscosity of a two component system.

We consider a gas of bosons and fermions as it will have applications to a gas of quarks and gluons or a gas of mesons and baryons.  
\beqa
\delta f_f(p)&=&-\nu_f n_p(1-n_p)\chi_f(p)\hat{p}^i \hat{p}^j\left< \partial_i u_j\right>\,, \nonumber\\
\delta f_b(p)&=&-\nu_b n_p(1+n_p)\chi_b(p)\hat{p}^i \hat{p}^j\left< \partial_i u_j\right> \, . 
\eeqa
The off-equilibrium correction $\chi$ takes the form
\beqa
\chi_b(p)&=&C_b(T)p^{2-\alpha_b} \, , \nonumber\\
\chi_f(p)&=&C_f(T)p^{2-\alpha_f} \,.
\eeqa
The goal is to find values of the coefficients $C_b$ and $C_f$ as a function of $T$ and $\eta/s$.

First we define the partial viscosity of each species
\beqa
\eta_f = \frac{\nu_f}{15}\int\frac{d^3p}{(2\pi)^3} p\chi_f(p)n_p\left[1-n_p\right]\,, \nonumber\\
\eta_b = \frac{\nu_b}{15}\int\frac{d^3p}{(2\pi)^3} p\chi_b(p)n_p\left[1+n_p\right] \,, 
\label{eq:D3}
\eeqa
which will yield a total viscosity of
\beqa
\eta=\eta_f+\eta_b\,.
\eeqa
For massive particles the phase space integrals must be done
numerically, but for massless particles, integrating \Eq{eq:D3} yields
\beqa
C_f(T) &=& \frac{7\pi^4\eta_f}
 {6 s_f T^{3-\alpha_f}\Gamma(6-\alpha_f)\zeta_-(5-\alpha_f)}\nonumber\, ,\\
C_b(T) &=& \frac{4\pi^4\eta_b}
 {3 s_b T^{3-\alpha_b}\Gamma(6-\alpha_b)\zeta(5-\alpha_b)}\, ,
\eeqa
with $\zeta_-(x) = \sum_{n=1}^{\infty} (-1)^{n-1} n^{-x} =
(1-2^{1-x}) \zeta(x)$.
Let us define $\mathcal{R}$ as the ratio of the partial viscosities,
\beqa
\frac{\eta_f}{\eta_b}\equiv \mathcal{R}\,.
\eeqa
Making use of the relations
\beqa
\eta&=&\eta_f+\eta_b=(1+\mathcal{R})\eta_b\,, \nonumber\\
s&=&s_f+s_b=v_f\frac{7\pi^2}{180}+v_b\frac{2\pi^2}{45} \,,
\eeqa
we find
\beqa
C_f(T) &=& \left(\frac{1+\frac{8\nu_b}{7\nu_f}}
   {1+\frac{1}{\mathcal{R}}}\right)\frac{\eta}{s}\times
 \frac{7\pi^4}{6 T^{3-\alpha_f}\Gamma(6-\alpha_f)
     \zeta_-(5-\alpha_f)} \,, \nonumber\\
C_b(T) &=& \left(\frac{1+\frac{7\nu_f}{8\nu_b}}
   {1+\mathcal{R}}\right)\frac{\eta}{s}\times
   \frac{4\pi^4}{3 T^{3-\alpha_b}\Gamma(6-\alpha_b)
   \zeta(5-\alpha_b)} \,.
\label{eq:2Cfinal}
\eeqa


\begin{thebibliography}{MM}

\bibitem{Adams:2005dq}
  J.~Adams {\it et al.}  [STAR Collaboration],
  Nucl.\ Phys.\  A {\bf 757}, 102 (2005)
  [arXiv:nucl-ex/0501009].
\bibitem{Adcox:2004mh}
  K.~Adcox {\it et al.}  [PHENIX Collaboration],
  Nucl.\ Phys.\  A {\bf 757}, 184 (2005)
  [arXiv:nucl-ex/0410003].
\bibitem{Back:2004je}
  B.~B.~Back {\it et al.},
  Nucl.\ Phys.\  A {\bf 757}, 28 (2005)
  [arXiv:nucl-ex/0410022].
\bibitem{Arsene:2004fa}
  I.~Arsene {\it et al.}  [BRAHMS Collaboration],
  Nucl.\ Phys.\  A {\bf 757}, 1 (2005)
  [arXiv:nucl-ex/0410020].

\bibitem{idealhydro}
  T.~Hirano and Y.~Nara,
  Nucl.\ Phys.\  A {\bf 743}, 305 (2004).
\, D.~Teaney, J.~Lauret and E.~V.~Shuryak,
Phys.\ Rev.\ Lett.\  {\bf 86}, 4783 (2001);  
{\it ibid} arXiv:nucl-th/0110037. 
\, P.~F.~Kolb, P.~Huovinen, U.~W.~Heinz and H.~Heiselberg,
Phys.\ Lett.\ B {\bf 500}, 232 (2001).
\, P.~Huovinen, P.~F.~Kolb, U.~W.~Heinz, P.~V.~Ruuskanen and S.~A.~Voloshin,
Phys.\ Lett.\ B {\bf 503}, 58 (2001).
\,  C.~Nonaka and S.~A.~Bass,
  Phys.\ Rev.\  C {\bf 75}, 014902 (2007). 

\bibitem{Abelev:2008ed}
B.~I.~Abelev {\it et al.} [STAR Collaboration],
Phys.\ Rev.\ C {\bf 77}, 054901 (2008) [arXiv:0801.3466 [nucl-ex]].

\bibitem{Adare:2006ti}
  A.~Adare {\it et al.}  [PHENIX Collaboration],
  Phys.\ Rev.\ Lett.\  {\bf 98}, 162301 (2007)
  [arXiv:nucl-ex/0608033].

\bibitem{coalesence}
  See for example the recent review of coalesence and constituent quark scaling:
P.~Sorensen,
  arXiv:0905.0174 [nucl-ex],  
  invited review for  {\it QGP4}, editors  R.~C.~Hwa and X.~N.~Wang.

\bibitem{Lin:2001zk}
  Z.~w.~Lin and C.~M.~Ko,
  Phys.\ Rev.\  C {\bf 65}, 034904 (2002)
  [arXiv:nucl-th/0108039].

\bibitem{Molnar:2003ff}
  D.~Molnar and S.~A.~Voloshin,
  Phys.\ Rev.\ Lett.\  {\bf 91}, 092301 (2003)
  [arXiv:nucl-th/0302014].

\bibitem{Greco:2003xt}
  V.~Greco, C.~M.~Ko and P.~Levai,
  Phys.\ Rev.\ Lett.\  {\bf 90}, 202302 (2003)
  [arXiv:nucl-th/0301093].

\bibitem{Fries:2003vb}
  R.~J.~Fries, B.~Muller, C.~Nonaka and S.~A.~Bass,
  Phys.\ Rev.\ Lett.\  {\bf 90}, 202303 (2003)
  [arXiv:nucl-th/0301087].
\bibitem{Baier:2006gy}
  R.~Baier and P.~Romatschke,
  Eur.\ Phys.\ J.\  C {\bf 51}, 677 (2007)
  [arXiv:nucl-th/0610108].

\bibitem{Romatschke:2007jx}
  P.~Romatschke,
  Eur.\ Phys.\ J.\  C {\bf 52}, 203 (2007)
  [arXiv:nucl-th/0701032].

\bibitem{Romatschke:2007mq}
  P.~Romatschke and U.~Romatschke,
  Phys.\ Rev.\ Lett.\  {\bf 99}, 172301 (2007)
  [arXiv:0706.1522 [nucl-th]].

\bibitem{Song:2007fn}
  H.~Song and U.~W.~Heinz,
  Phys.\ Lett.\  B {\bf 658}, 279 (2008)
  [arXiv:0709.0742 [nucl-th]].
\bibitem{Dusling:2007gi}
  K.~Dusling and D.~Teaney,
  Phys.\ Rev.\  C {\bf 77}, 034905 (2008)
  [arXiv:0710.5932 [nucl-th]].
\bibitem{Huovinen:2008te}
  P.~Huovinen and D.~Molnar,
  Phys.\ Rev.\  C {\bf 79}, 014906 (2009)
  [arXiv:0808.0953 [nucl-th]].

\bibitem{Song:2007ux}
  H.~Song and U.~W.~Heinz,
  Phys.\ Rev.\  C {\bf 77}, 064901 (2008)
  [arXiv:0712.3715 [nucl-th]].
\bibitem{Bozek:2007qt}
  P.~Bozek,
  Phys.\ Rev.\  C {\bf 77}, 034911 (2008)
  [arXiv:0712.3498 [nucl-th]].
\bibitem{Hiscock:1983zz}
  W.~A.~Hiscock and L.~Lindblom,
  Annals Phys.\  {\bf 151}, 466 (1983).
\bibitem{Hiscock:1985zz}
  W.~A.~Hiscock and L.~Lindblom,
  Phys.\ Rev.\  D {\bf 31}, 725 (1985).
\bibitem{IS}
W. Israel, Ann. Phys. {\bf 100} (1976) 310; W. Israel and J.M. Stewart, Phys. Lett. {\bf 58A} (1976) 213.
\bibitem{OG}
M. Grmela, H.C. \"{O}ttinger, Phys. Rev. E {\bf 56}, 6620 (1997). H.C. \"{O}ttinger, M. Grmela, Phys. Rev. E {\bf 56}, 6633 (1997). H.C. \"{O}ttinger, Phys. Rev. E {\bf 57}, 1416 (1993).
\bibitem{Ottinger}
H. C. \"{O}ttinger, Physica A {\bf 254} (1998) 433-450.
\bibitem{CF}
F. Cooper and G. Frye, Phys. Rev. D. {\bf 10}, 186 (1974).
 
\bibitem{Teaney:2009qa}
  See for example, D.~A.~Teaney,
  arXiv:0905.2433 [nucl-th],
  invited review for  {\it QGP4}, editors  R.~C.~Hwa and X.~N.~Wang.

\bibitem{Jeon:1994if}
  S.~Jeon,
  Phys.\ Rev.\  D {\bf 52}, 3591 (1995)
  [arXiv:hep-ph/9409250].

\bibitem{Arnold:2001ba}
  P.~Arnold, G.~D.~Moore and L.~G.~Yaffe,
  JHEP {\bf 0111}, 057 (2001)
  [arXiv:hep-ph/0109064].
\bibitem{Arnold:2002zm}
  P.~Arnold, G.~D.~Moore and L.~G.~Yaffe,
  JHEP {\bf 0301}, 030 (2003)
  [arXiv:hep-ph/0209353].
\bibitem{Arnold:2000dr}
  P.~Arnold, G.~D.~Moore and L.~G.~Yaffe,
  JHEP {\bf 0011}, 001 (2000)
  [arXiv:hep-ph/0010177].
\bibitem{Juhee}
 J.~Hong and D.~Teaney, in preparation.

\bibitem{Thoma:1992kq}
  M.~H.~Thoma and M.~Gyulassy,
  Nucl.\ Phys.\  A {\bf 544} (1992) 573C.

\bibitem{Braaten:1991jj}
  E.~Braaten and M.~H.~Thoma,
  Phys.\ Rev.\  D {\bf 44}, 1298 (1991); {\it ibid.}
  Phys.\ Rev.\ D {\bf 44}, 2625 (1991).

\bibitem{Arnold:2003zc}
  P.~Arnold, G.~D.~Moore and L.~G.~Yaffe,
  JHEP {\bf 0305}, 051 (2003)
  [arXiv:hep-ph/0302165].

\bibitem{Bjorken:1982tu}
  J.~D.~Bjorken,
  FERMILAB-PUB-82-059-THY.

\bibitem{Baym:1990uj}
  G.~Baym, H.~Monien, C.~J.~Pethick and D.~G.~Ravenhall,
  Phys.\ Rev.\ Lett.\  {\bf 64}, 1867 (1990).
\bibitem{Heiselberg:1994vy}
  H.~Heiselberg,
  Phys.\ Rev.\  D {\bf 49}, 4739 (1994)
  [arXiv:hep-ph/9401309].
\bibitem{Baier:1996kr}
  R.~Baier, Y.~L.~Dokshitzer, A.~H.~Mueller, S.~Peigne and D.~Schiff,
  Nucl.\ Phys.\  B {\bf 483}, 291 (1997)
  [arXiv:hep-ph/9607355].
\bibitem{Baier:1996sk}
  R.~Baier, Y.~L.~Dokshitzer, A.~H.~Mueller, S.~Peigne and D.~Schiff,
  Nucl.\ Phys.\  B {\bf 484}, 265 (1997)
  [arXiv:hep-ph/9608322].
\bibitem{Arnold:2002ja}
  P.~Arnold, G.~D.~Moore and L.~G.~Yaffe,
  JHEP {\bf 0206}, 030 (2002)
  [arXiv:hep-ph/0204343].
\bibitem{Arnold:2008zu}
  P.~Arnold and C.~Dogan,
  Phys.\ Rev.\  D {\bf 78}, 065008 (2008)
  [arXiv:0804.3359 [hep-ph]].
\bibitem{Bass:2008rv}
  S.~A.~Bass, C.~Gale, A.~Majumder, C.~Nonaka, G.~Y.~Qin, T.~Renk and J.~Ruppert,
  Phys.\ Rev.\  C {\bf 79}, 024901 (2009)
  [arXiv:0808.0908 [nucl-th]].

\bibitem{HT}
  X.~f.~Guo and X.~N.~Wang,
  Phys.\ Rev.\ Lett.\  {\bf 85}, 3591 (2000).
\,  X.~N.~Wang and X.~f.~Guo,
  Nucl.\ Phys.\  A {\bf 696}, 788 (2001).
\,  B.~W.~Zhang and X.~N.~Wang,
  Nucl.\ Phys.\  A {\bf 720}, 429 (2003).
\,  A.~Majumder, E.~Wang and X.~N.~Wang,
  Phys.\ Rev.\ Lett.\  {\bf 99}, 152301 (2007).
\,  A.~Majumder and B.~Muller,
  Phys.\ Rev.\  C {\bf 77}, 054903 (2008).
\,  A.~Majumder, R.~J.~Fries and B.~Muller,
  Phys.\ Rev.\  C {\bf 77}, 065209 (2008).

\bibitem{AMYeloss}
  S.~Jeon and G.~D.~Moore,
  Phys.\ Rev.\  C {\bf 71}, 034901 (2005)


\bibitem{ASW}
B. G. Zakharov, JETP Lett. {\bf 63}, 952 (1996), hep-ph/9607440.  B. G. Zakharov, JETP Lett. {\bf 65}, 615 (1997), hep-ph/9704255.  B. G. Zakharov, Phys. Atom. Nucl. {\bf 61}, 838 (1998), hep-ph/9807540.  C. A. Salgado and U. A. Wiedemann, Phys. Rev. {\bf D68}, 014008 (2003), hep-ph/0302184.  U. A. Wiedemann, Nucl. Phys. {\bf B582}, 409 (2000), hep-ph/0003021.  U. A. Wiedemann, Nucl. Phys. {\bf B588}, 303 (2000), hep-ph/0005129.  C. A. Salgado and U. A. Wiedemann, Phys. Rev. Lett.  {\bf 89}, 092303 (2002), hep-ph/0204221. N. Armesto, C. A. Salgado, and U. A.Wiedemann, Phys. Rev. Lett. {\bf 94}, 022002 (2005), hep-ph/0407018.

\bibitem{Laine:2006cp}
  M.~Laine and Y.~Schroder,
  Phys.\ Rev.\  D {\bf 73}, 085009 (2006)
  [arXiv:hep-ph/0603048].
\bibitem{PHENIXv2a}
Arkadij Taranenko  for the PHENIX Collaboration, 
presented at \emph{Quark Matter 2009}, 
Knoxville, Tennessee, 
March 30--April 4 (2009).

\bibitem{PHENIXv2b}
S.~Huang  [PHENIX Collaboration],
J.\ Phys.\ G {\bf 36} (2009) 064061.

\bibitem{KSH}
P. F. Kolb, J. Sollfrank, and U. Heinz, Phys. Rev. C {\bf 65}, 054909 (2000).
\bibitem{Ollitrault:1992bk}
  J.~Y.~Ollitrault,
  Phys.\ Rev.\  D {\bf 46}, 229 (1992).

\bibitem{AQM}
  E.~M.~Levin and L.~L.~Frankfurt, Pisma ZhETP, {\bf 3}, 105 (1965). 
  H.~J.~Lipkin and F.~Sheck, Phys.\ Rev.\ Lett.\ {\bf 16}, 71 (1966).

\bibitem{Bleicher:2000sx}
  M.~Bleicher and H.~Stoecker,
  Phys.\ Lett.\  B {\bf 526}, 309 (2002)
  [arXiv:hep-ph/0006147].

\bibitem{Dusling:2009zz}
  K.~Dusling,
  Acta Phys.\ Polon.\  B {\bf 40}, 963 (2009).

\end{thebibliography}
\end{document}